\documentclass[11pt,a4paper,epsfig]{article}
\usepackage{graphicx}

\def\title#1{\begin{centering}\Large\bf #1 \\[5mm]\end{centering}}
\def\author#1{\begin{centering}#1 \\[2mm]\end{centering}}
\def\address#1{\begin{centering}\em #1 \\[5mm]\end{centering}}
\def\pacs#1{\em #1 \\[5mm]}
\def\submitto#1{\em #1 \\[5mm]}

\usepackage{amsmath}
\usepackage{gensymb}
\usepackage{wasysym}
\usepackage{upgreek}
\usepackage[english]{babel}

\usepackage{textcomp}
\usepackage{subfig}
\usepackage{caption}

\usepackage{float}
\DeclareGraphicsExtensions{,.ps,.gz,.ps.gz}
\usepackage{epstopdf}

\usepackage[nottoc, notlof, notlot]{tocbibind}

\begin{document}
\title{Heterodyne non-demolition measurements on cold atomic samples: towards the preparation of non-classical states for atom interferometry.}
\author{S. Bernon$^1$, T. Vanderbruggen$^1$, R. Kohlhaas$^1$, A. Bertoldi$^1$, A. Landragin$^2$ and P. Bouyer$^{1,3}$.}
\address{$^1$ Laboratoire Charles Fabry de l'Institut d'Optique, CNRS and Univ. Paris-Sud\\ Campus Polytechnique, RD 128, F-91127 Palaiseau cedex, France}
\address{$^2$ LNE-SYRTE, Observatoire de Paris, CNRS and UPMC\\ 61 avenue de l'Observatoire, F-75014 Paris, France}
\address{$^3$ Laboratoire Photonique, Numerique et Nanosciences - LP2N\\ Universite Bordeaux - IOGS - CNRS : UMR 5298\\ Bat A30, 351 cours de la liberation, Talence, France}

\begin{abstract}
We report on a novel experiment to generate non-classical atomic states  \textit{via} quantum non-demolition (QND) measurements on cold atomic samples prepared in a high finesse ring cavity. The heterodyne technique developed for the QND detection exhibits an optical shot-noise limited behavior for local oscillator optical power of a few hundred $\upmu$W, and a detection bandwidth of several GHz. This detection tool is used in single pass to follow non destructively the internal state evolution of an atomic sample when subjected to Rabi oscillations or a spin-echo interferometric sequence.
\end{abstract}

%    37.10.De Atom cooling methods
%
%    37.10.Vz Mechanical effects of light on atoms, molecules, and ions
%
%    32.60.+i Zeeman and Stark effects
%
%    42.50.Hz Strong-field excitation of optical transitions in quantum systems; multiphoton processes; dynamic Stark shift

\submitto{Published in New J. Phys. \textbf{13} 065021 (2011)}
%\date{\no}
%\maketitle
\newpage

\section{Introduction}

Quantum metrology \cite{lee02} is the field of quantum information that studies the ultimate amount of information
obtainable when measuring a given observable of a system. Quantum state engineering techniques are thus developed
to maximize the sensitivity of measurement devices. 

In quantum optics, after the first observation of squeezed states \cite{slusher85}, squeezing enhancing techniques
have been strongly investigated, leading to very highly squeezed states \cite{vahlbruch08} and NOON states
\cite{afek10}. The benchtest to study the practical application of quantum technology is gravitational wave
detection: the instruments used nowadays are reaching classical limit due to the optical radiation pressure on the mirrors. Hence, further sensitivity improvements require quantum engineering techniques. Recently, squeezing
enhancement has been proven in a gravitational wave detector prototype \cite{goda08}.

State-of-the-art atomic detectors such as clocks \cite{santarelli99}, inertial sensors
\cite{Gauguet2009}, and magnetometers \cite{wasilewski10} have also reached the standard quantum limit given by the
atomic shot-noise. To go further, the sensitivity can be enhanced either increasing the number of atoms, or reducing the effect of quantum noise. In this context, atomic spin-squeezed states \cite{Kitagawa1993} have been
recently achieved \cite{Schleier2010,Louchet2010,Esteve2008,Gross2010,Riedel2010}, and they allowed sensitivity enhancement in atomic clocks \cite{Louchet2010,Wineland1992}. 

In this paper we present an apparatus designed to generate cold atomic samples in an optical cavity and to perform heterodyne non-demolition measurement in the perspective of spin-squeezed states preparation. At first an overview of the experimental apparatus and a characterization of the high-finesse optical cavity are presented. The second part is 
dedicated to the heterodyne nondestructive measurement of an atomic population. After  a presentation of the experimental set-up, we demonstrate the  non-demolition measurement of Rabi oscillations and characterize the influence of the probe on the oscillations. We finally show the real-time measurement of the internal atomic state evolution in a Ramsey interferometer.

\section{The experimental apparatus}

\begin{figure}
\center
\includegraphics[width= 14 cm]{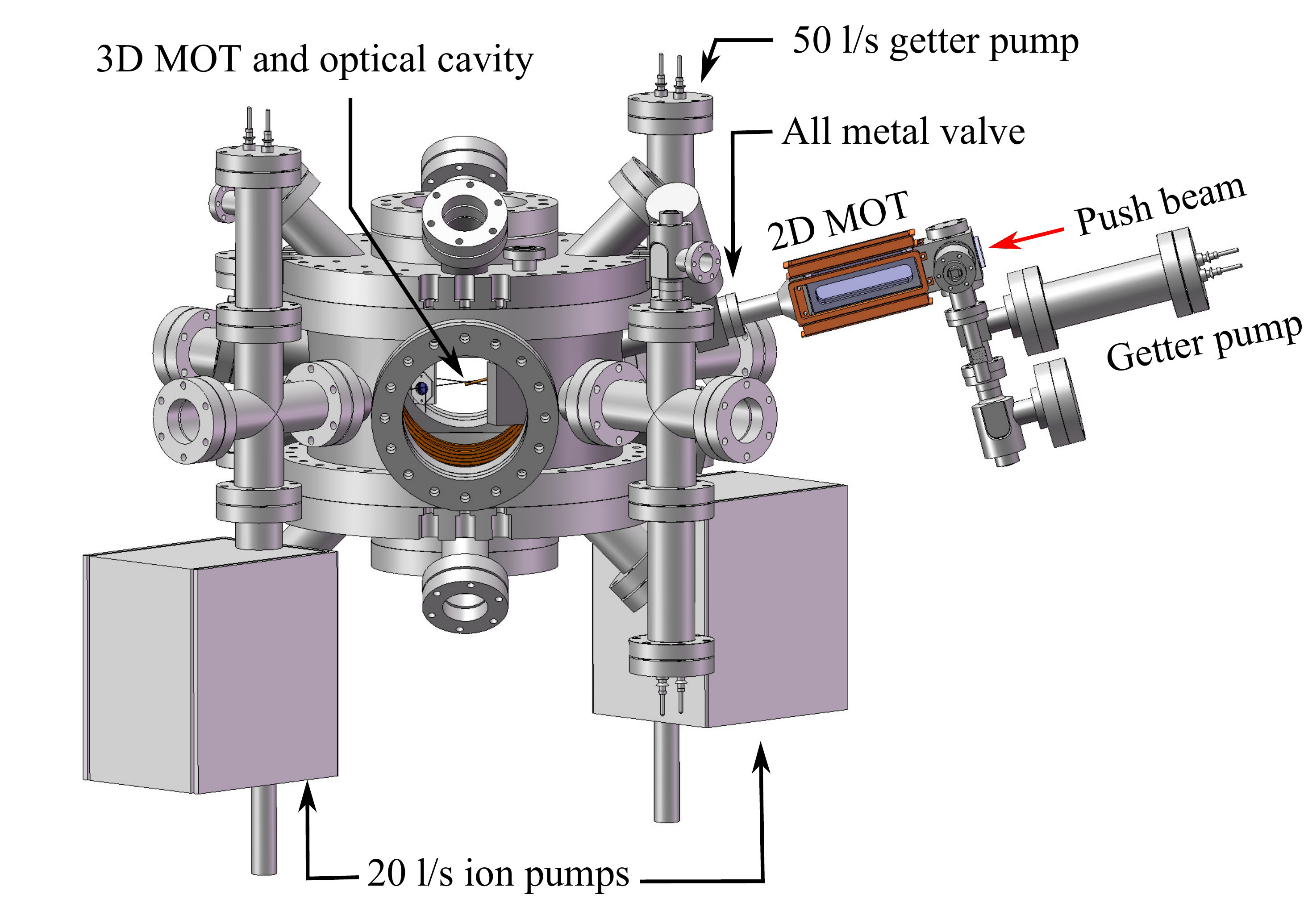}
\caption{Overview of the vacuum system showing the compact 2D MOT on the right and the science chamber with the high finesse cavity on the left.}
\label{fig:general_view}
\end{figure}

The apparatus is composed of two chambers: in the first a 2D MOT is operated, and used as a source of cold atoms; the second one is the science chamber where the 3D MOT and the experiments are performed in the center of a crossed optical cavity (figure \ref{fig:general_view}).

\subsection{2D and 3D MOTs}

The 2D MOT vacuum chamber is a titanium machined piece with glue sealed windows, which are antireflection coated at $780$ nm on both sides. The science chamber is made of non magnetic stainless steel (AISI type 316LN) and titanium; its viewports are antireflection coated at $780$ nm. To operate the resonator enhanced optical trap, the viewports on the horizontal plane are additionally antireflection coated at 1560 nm.

The pressure in the 2D MOT chamber is below $7\times10^{-8}$ mbar, whereas in the science chamber it is below $10^{-9}$ mbar. The differential vacuum operation relies on a reduced conductivity between the two chambers.

%To reach an ultra high vacuum the whole assembly was baked at high temperature, except the cavity and the MOT coils, which were treated separately. This procedure was adopted to prevent contaminating the science chamber with the kapton insulation of the coils.

\begin{figure}
\center
\begin{minipage}[c]{0.45\textwidth}
\begin{center}
\subfloat[][]{
\includegraphics[width= 7 cm]{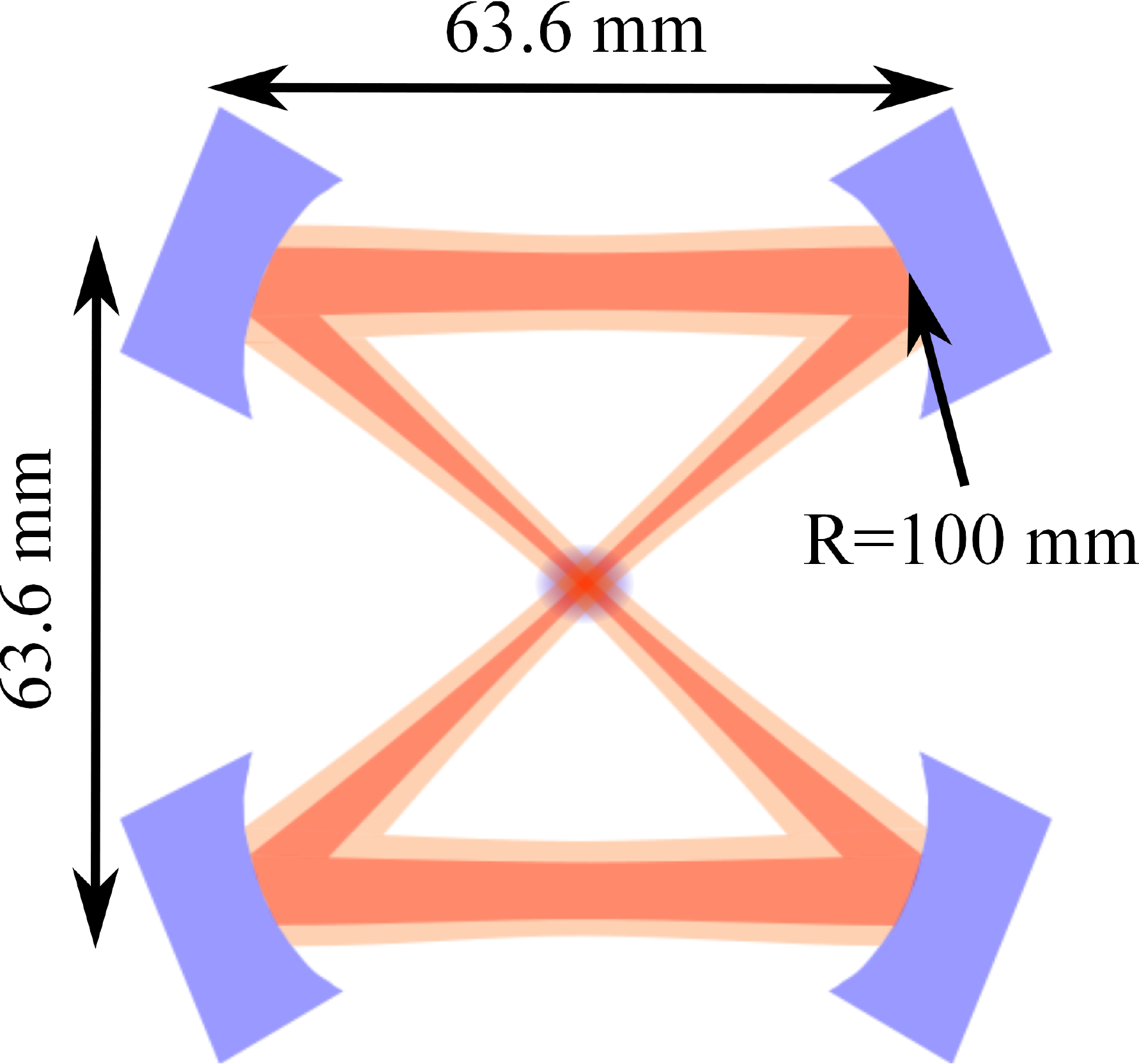}}
\end{center}
\end{minipage}
\vspace{0.5 cm}
\begin{minipage}[c]{0.45\textwidth}
\begin{center}
\subfloat[][]{
\includegraphics[width= 7.5 cm]{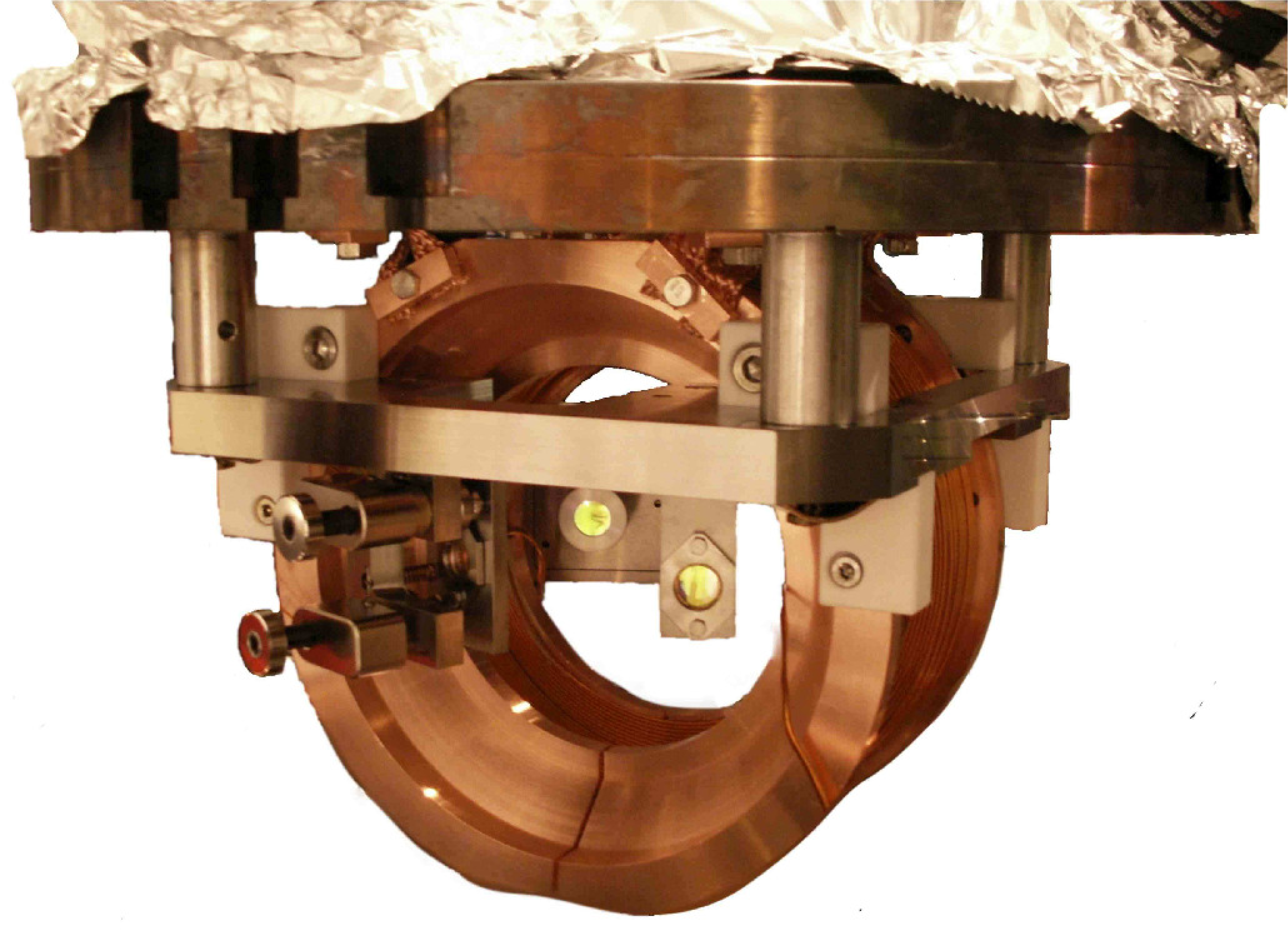}
\label{fig:cavity_clean}}
\end{center}
\end{minipage}
\caption{The folded optical cavity. (a) Schematic of the cavity geometry and (b) the inner vacuum set-up with the suspended cavity and the coils generating the MOT magnetic field.}
\label{fig:cavity}
\end{figure}

The 2D MOT produces a bright source of pre-cooled atoms used to load the 3D MOT. The rubidium background
pressure is obtained from the vapor pressure of a 1 g rubidium sample. The loading rate
of the 3D MOT by the 2D MOT is about $5\times 10^8$ at.$s^{-1}$ at room temperature. The MOT lifetime
is $\sim$ 20 s. The center of the 3D MOT is placed $3$ mm above the 2D MOT jet to avoid direct collisions with
thermal atoms, and is centered on the cavity crossing region where the atoms can be optically trapped in the
resonator. The 3D MOT coils are mounted in vacuum and are mechanically tightened using ceramic supports to the
titanium plate holding the cavity (figure \ref{fig:cavity}(b)). The heat produced by the 3D MOT coils ($2\times9.6$ W)
is directly dissipated towards the vacuum chamber through copper braided links. These coils create a $2.2\
\rm{G\ cm}^{-1}\rm{A}^{-1}$ gradient and are normally operated at $4$ A.

The MOT beams are generated by external cavity diode lasers \cite{Baillard2006} amplified to $1~$W by tapered amplifiers. Fiber coupled splitters (Sch\"{a}fter+Kirchhoff and OzOptics) are adopted to obtain the independent MOT beams in a compact and reliable design.

The repump light frequency is fixed on the $\left|F=1\right\rangle\rightarrow \left|F=2\right\rangle$ atomic transition while the cooling and probing light are frequency locked to the repump laser. They can be continuously frequency shifted over a range of 400 MHz. 

\subsection{The high finesse cavity}
\label{high_finess}

The optical cavity, that is resonant for both 1560 nm and 780 nm,  is of dual interest in this experiment. In the first instance, it allows to produce a deep optical potential to trap atoms using a low power laser source by exploiting the build-up effect of the resonator at 1560 nm. On the other hand, it enhances the light-atom coupling thus improving the non-demolition measurement efficiency realized at 780 nm.

\subsubsection{Mechanical design}
The cavity, shown in figure \ref{fig:cavity}, is made up of four mirrors placed at the corners of a square with a
diagonal of $90$ mm. All the mechanical components of the cavity are realized in titanium to maintain a low
magnetic environment for the trapped atoms. The size of the cavity results from a compromise to maximize
the optical access while minimizing the cavity mode to ensure a tight trapping. The square geometry
enables a $90\degree$ crossing angle which avoids an interference pattern in the crossing region once the
polarization is set to be in the cavity plane. The mirrors are all identical and are plane-concave with a radius of curvature of
$R=100$ mm.  They have a diameter of $1/2$ inch. All the parts are tightly mounted on a $15$ mm thick titanium
plate, which gives its mechanical stability to the set-up. Two mirror mounts are completely fixed. One mount is actuated under vacuum by piezoelectric actuators (Newfocus Picomotors) which
allow for a coarse alignment of the cavity. The last mount is a piezo-actuated three-axis nano-positioning
system (Madcitylabs M3Z) with maximal angular displacement of 2 mrad and translation of $50~\upmu$m. It is used
to finely adjust the cavity crossing angle and dynamically control the cavity length.

\subsubsection{Spectral properties}
The four mirrors of the cavity are highly reflecting at both 1560 nm and 780 nm.
The free spectral range (FSR) of the cavity is $976.2$ MHz and the cavity FWHM linewidth is $\gamma=546$ kHz at $1560$ nm. 
The FSR is measured by simultaneously injecting the carrier and a sideband created by an electro-optic modulator (EOM) in two adjacent longitudinal modes. $\gamma$ is obtained from the Lorentzian profile of the transmission when the sideband is scanned across the resonance.
The finesse is then $\mathcal{F}_{1560}=\mathrm{FSR}/\gamma=1788$. The corresponding mirror amplitude reflectivity obtained from the formula $\mathcal{F}=\pi r^2/(1-r^4)$, is $r=99.956\%$ at 1560 nm. At 780 nm, the reflectivity measured by the manufacturer is $r=99.99923\%$, \text{i.e.} an expected finesse $\mathcal{F}_{780}=102000$.

\subsubsection{Geometrical characterization}
To estimate the mode parameters, the resonator has been analyzed with the ABCD matrix formalism \cite{kogelnik66}. In more detail, the cavity is converted into an equivalent sequence of lenses and self consistency is postulated.
 In the adopted geometry, an astigmatism arises from the nonzero angle of incidence $\theta$ on the curved mirrors. The effective radii of curvature are then $R_{\parallel}=R\cos\theta$  in the plane defined by the cavity (horizontal plane) and $R_{\bot}= R/\cos{\theta}$ in the orthogonal plane.
\begin{figure}
\center
\includegraphics[width= 0.9\textwidth]{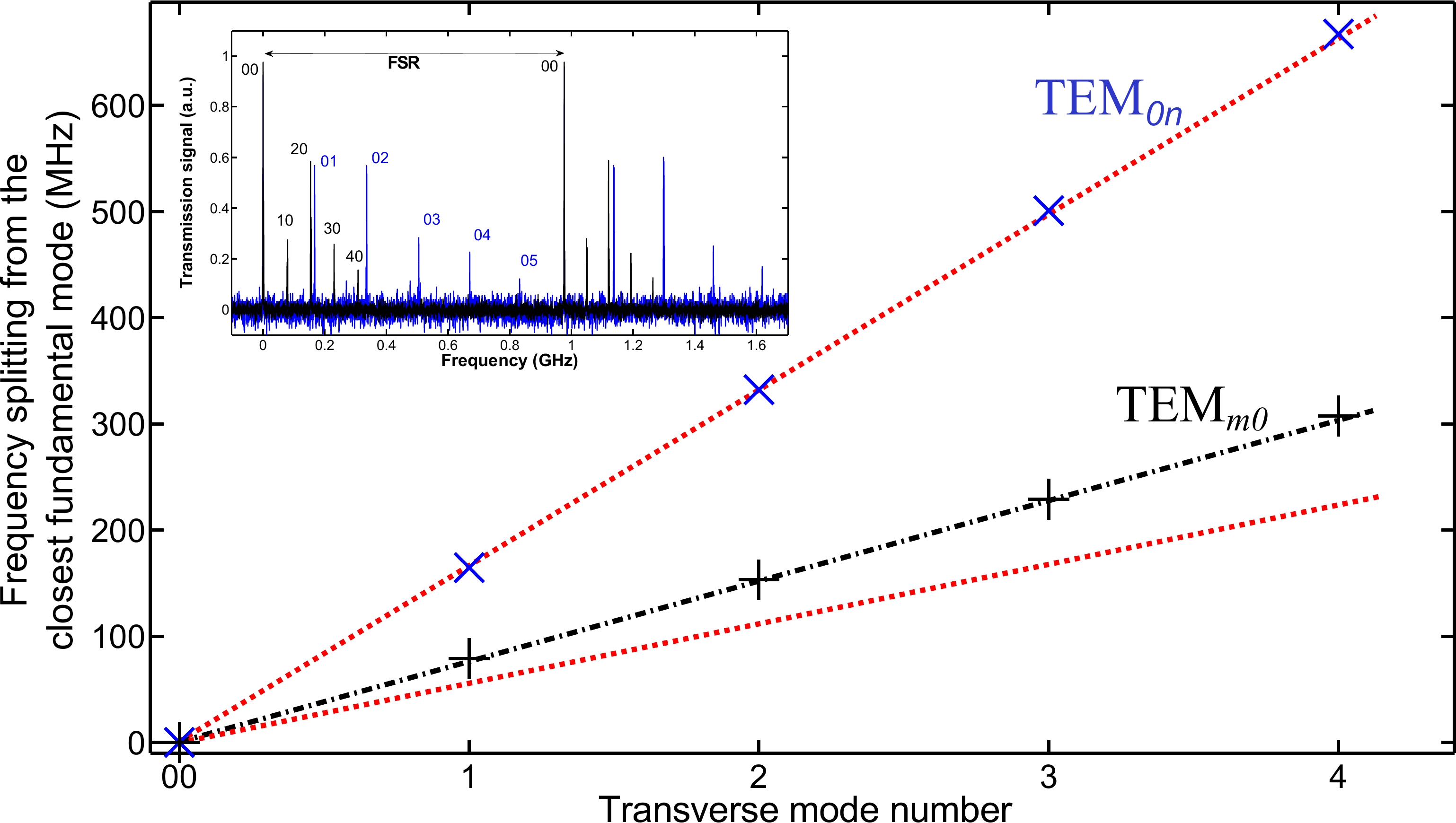}
\caption{Frequency splitting of the transverse mode relatively to the closest fundamental mode. The red dashed lines represent equation (\ref{eq:transverse}) for the two orthogonal directions when only the off axis incidence on the mirrors is taken into account. For the horizontal direction, the black dashed-dotted line includes also the corrections on the  effective radius of curvature. In inset, the cavity transmission signal when the cavity length is scanned for a horizontal (black, $\textrm{TEM}_{m0}$) and a vertical (blue, $\textrm{TEM}_{0n}$) misalignment of the injection beam.}
\label{fig:Transverse_mode_splitting}
\end{figure}

We measured the geometrical properties of the cavity by analyzing the transverse mode spectrum at 1560 nm. Since the cavity is astigmatic, the transverse modes have Hermite-Gauss profiles. The horizontal mode inter-spacing is 78.9 MHz and the vertical one is 164.6 MHz. 
The mode eigenfrequencies are related to the Gouy phase \cite{Gouy1890} acquired by the photons while traversing the beam waist and are thus linked to the cavity geometry. The absolute mode frequencies are given by \cite{Siegman1986}:
\begin{equation}
\nu_{p,m,n}=\frac{c}{2\pi L}\left[2 p \pi + \left(m+\frac{1}{2} \right) \textrm{Arg}\left(A_m+\frac{B_m}{q_m}\right) + \left(n+\frac{1}{2} \right) \textrm{Arg}\left(A_n+\frac{B_n}{q_n}\right) \right],
\label{eq:transverse}
\end{equation}
where $p$ is the longitudinal mode order, $m$ and $n$ are the horizontal ($\textrm{TEM}_{m0}$) and vertical ($\textrm{TEM}_{0n}$) transverse mode numbers, $A_m$ and $B_m$ factors refer to the ABCD matrix element in the $m$ direction, and $q_m$ is the complex radius of curvature of the beam. The parameters $A$, $B$ and $q$ contain the geometrical information of the cavity (length, angle and radius of curvature of the mirrors).

Figure \ref{fig:Transverse_mode_splitting} compares the experimental measurement of the transverse mode inter-spacing and its \textit{a priori} calculation in red dashed line. The discrepancy of the horizontal splitting is corrected by a phenomenological factor $\alpha_{\parallel}$ that, for example, accounts for aberrations. This factor is applied on the horizontal effective radius of curvature $R_{\parallel}=\alpha_{\parallel} R\cos\theta$, and is adjusted on the slope of the data giving $\alpha_{\parallel}=1.020(5)$ (black curve in figure \ref{fig:Transverse_mode_splitting}). In the adopted quasi-concentric configuration, the parallel direction is the closest one to the instability regime. A $2\%$ deviation of $\alpha_{\parallel}$ induces a noticeable frequency shift as it can be seen in figure \ref{fig:Transverse_mode_splitting}.
Including this correction for the astigmatism, we infer a horizontal waist $\omega_{\parallel}=93.1~\upmu$m to be compared with $98(1)~\upmu$m found using a tomographic measurement \cite{brantut08,bertoldi10} and reported in section \ref{sec:tomography}. The vertical waist is calculated to be $\omega_{\bot}=129.8~\upmu$m. The corresponding Rayleigh ranges are $z_{r}^{\parallel}=17.46~$mm and $z_{r}^{\bot}=33.9~$mm.

\subsubsection{Frequency lock of the 1560 nm laser to the cavity}
The optical cavity is pumped with the radiation produced by a single longitudinal mode distributed feedback erbium doped fiber laser (DF-EDFL) near 1560 nm (Koheras laser from NKT Photonics), amplified with a 5 W erbium doped fibered amplifier (EDFA, from Keopsys). The typical laser linewidth is of the order of a few kHz, its output power is 100 mW single mode and is linearly polarized. The radiation is coupled to the resonator through a beam expander and a tilted doublet, whose angle allows us to mode-match the astigmatism.  
The laser is locked to a transversal cavity mode using the Pound-Drever-Hall technique \cite{drever83}. A low frequency feedback is applied on the piezoelectric element controlling the laser cavity length. 
The fast frequency correction is realized with an acousto-optic modulator in double-pass configuration with  a bandwidth of 250 kHz. More details about the implemented lock scheme and its noise performance can be found in \cite{bertoldi10}.

\subsubsection{Light-shift tomography of the cavity optical potential}
\label{sec:tomography}
The atoms released from the MOT, operated in the crossing region of the two cavity arms, are used to characterize \textit{in-situ} the resonator geometry. The light at 1560 nm that pumps the cavity is close to the rubidium transitions 5P$_{3/2}$--4D$_{3/2,5/2}$ at 1529~nm. This causes a strong differential light shift on the two levels of the D$_2$ transition. More precisely, at 1560 nm the scalar polarizability of the 5P$_{3/2}$ level is 47.7 times that of the 5S$_{1/2}$ level. As a consequence, by changing the detuning of the probe beam with regard to the D$_2$ transition it is possible to bring into resonance atoms placed at different positions in the optical potential \cite{brantut08}. 

\begin{figure}
\centering
\includegraphics[width=1\textwidth]{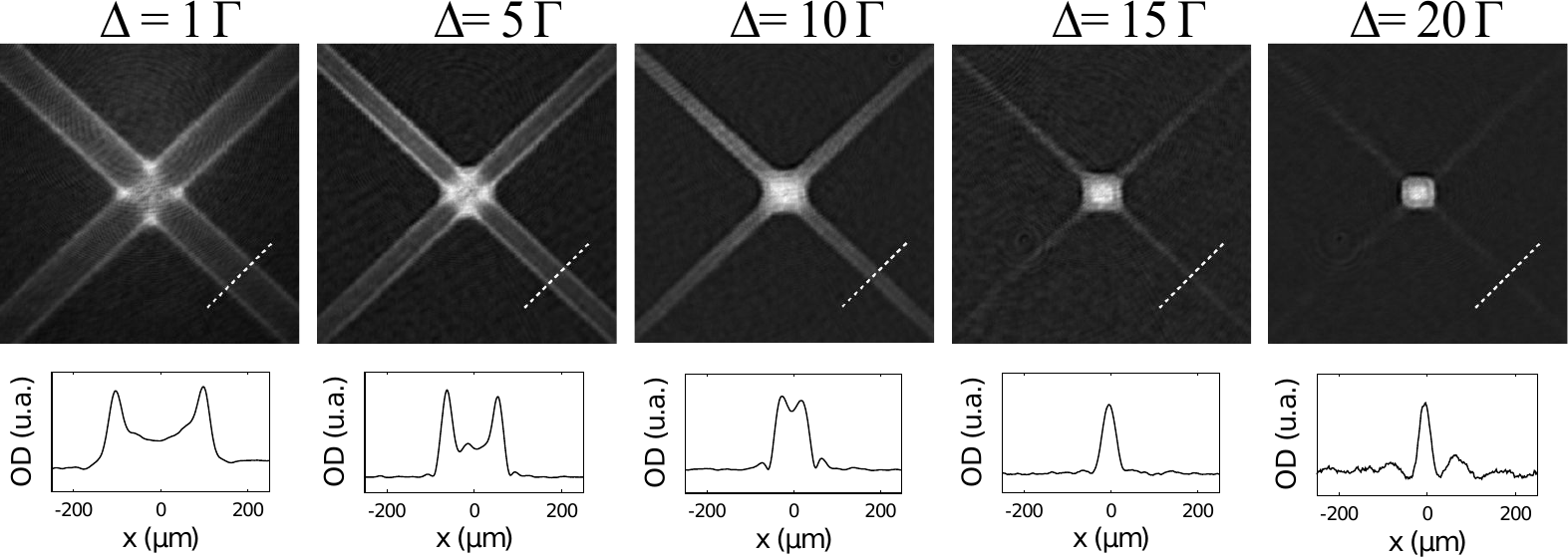}
\caption{\label{fig:tomographyTEM00}(Top) Tomographic images of the optical potential setting the probe frequency to different detuning with respect to the D$_2$ line. (Bottom) Integral optical density obtained by projecting the upper images on the 45$\degree$ dashed line crossing one arm of the cavity.}
\end{figure}

\begin{figure}
\centering
\includegraphics[width=0.6\textwidth]{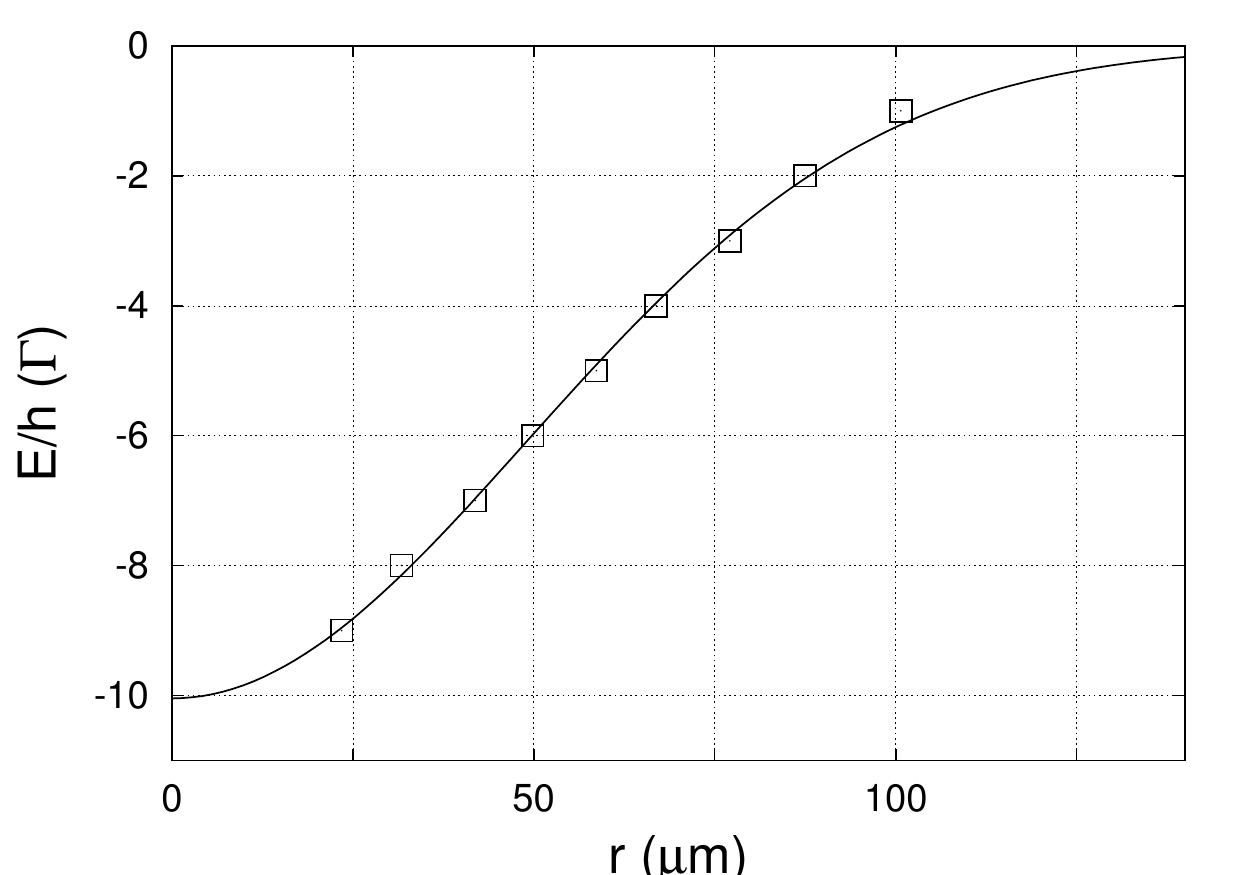}
\caption{\label{fig:tem00profile}Optical potential depth when the 1560 nm laser is locked to the fundamental transversal mode of the optical resonator. The measurement was obtained by projecting the signal relative to one arm of the cavity. The result of the fit with the first Hermite-Gauss mode is shown with a solid line.}
\end{figure}

This method provides the isopotential lines of the crossed dipole trap, as shown in figure \ref{fig:tomographyTEM00}, where the 1560 nm laser is locked to the fundamental transversal mode (TEM$_{00}$). In each shot, the probe detuned by $\delta$ is absorbed by the atoms with potential energy $U(\textbf{r})=-\epsilon_0 c \textrm{Re}(\alpha) I(\textbf{r})/2=\hbar \delta/(47.7-1)$ in the ground level; for this atomic class the probe is shifted into resonance because of the local differential light shift. The optical density was projected on an axis parallel to one cavity arm (lower row in figure \ref{fig:tomographyTEM00}), and the profile was fitted to obtain the distance between the couples of isopotential lines. Plotting the position of these isolines versus the probe detuning provides the cavity mode profile, as shown in figure \ref{fig:tem00profile}. A fit with the first Hermite-Gauss mode gives a waist $\omega_{\parallel} = 98(1)~\mu m$. 
In \cite{bertoldi10} a similar analysis was performed also for the TEM$_{10}$ and TEM$_{20}$ modes, obtaining consistent results.

\subsubsection{Scattering and losses}
The differential light shift presented in section \ref{sec:tomography} is proportional to the intracavity intensity and to the atomic polarizability ($\alpha=6.83\ 10^{-39}$ Jm$^2$V$^{-2}$ \cite{Arora2007,Safronova2006}). In a four mirror geometry, the intracavity power $P_{\rm intra}$ is related to the power at one output $P_{\rm out}$ by $P_{\rm intra}= 2(1+x) \mathcal{F} P_{\rm out}/\pi$, where $x=\kappa_L^2/t^2$ represents the loss  of the mirrors  $\kappa_L$ in units of the mirror transmission $t$. Combining  the measurement of the differential light shift caused by the intracavity power with the output power in one arm, we infer scattering losses on the mirrors that are $1.5~$times the mirror transmission.

Considering these extra-losses, the maximum coupling efficiency achievable is $36\%$. The maximum coupling obtained experimentally is $35(2)\%$.
The backward scattering on the cavity mirrors is measured to be $1.9\permil$ in intensity compared to the forward direction.  This light, that propagates in the reverse direction, interferes with the forward propagating light and causes a $4.38\%$ modulation depth of the optical potential. The standing wave formed by the retro-reflection is phase related to the mirror positions and is then sensitive to thermal drift and acoustic noise affecting the mirrors.

\section{Trapping in a folded cavity}
The optical resonator is used to enhance the 1560 nm input intensity to obtain a high intracavity power suitable to realize a far off resonance optical dipole trap for neutral atoms. More precisely, with an incident power of 5 W an intracavity power of $200~W$ is reached. Moreover, at the cavity crossing, the potential presents high trapping frequencies in all spatial directions. At full power, a trap depth of $U_0=1.4$ mK is obtained and the trapping frequencies are $\omega_{x} / 2 \pi = \omega_{y} / 2 \pi = 1.2$ kHz and $\omega_{z} / 2 \pi = 1.6$ kHz.

The dipole trap is loaded from a MOT in a similar way as described in \cite{clement09}. For the loading, the depth of the FORT starts at $\sim100~\upmu$K. A compressed MOT phase is operated by detuning the MOT beams to 5 $\Gamma$ from the atomic resonance for 50 ms. The cloud is compressed and the atomic density increases in the FORT region. The cooling beams are then further detuned to 40 $\Gamma$ for 100 ms. The high detuning has been chosen to red-detune the light with respect to the atomic transition even in the presence of the strong light-shift. Moreover, in the FORT region the repump light is out of resonance and the atoms are accumulated in the F=1 hyperfine state, realizing in this way a dark MOT induced by the atomic light-shift. Finally, the cooling radiation is turned off and the power in the FORT is ramped up to full power in 10 ms. 

At the end of the loading sequence about 20 million atoms are loaded at the crossing region of the dipole trap. After thermalization, the sample temperature is $230~\upmu$K, leading to a temperature to trap depth ratio $k_B T/U_0 \sim 6$. The trapping time is limited to 6.6 s by one body losses.

\begin{figure}[!h]
\begin{center}
\includegraphics[width=12cm,keepaspectratio]{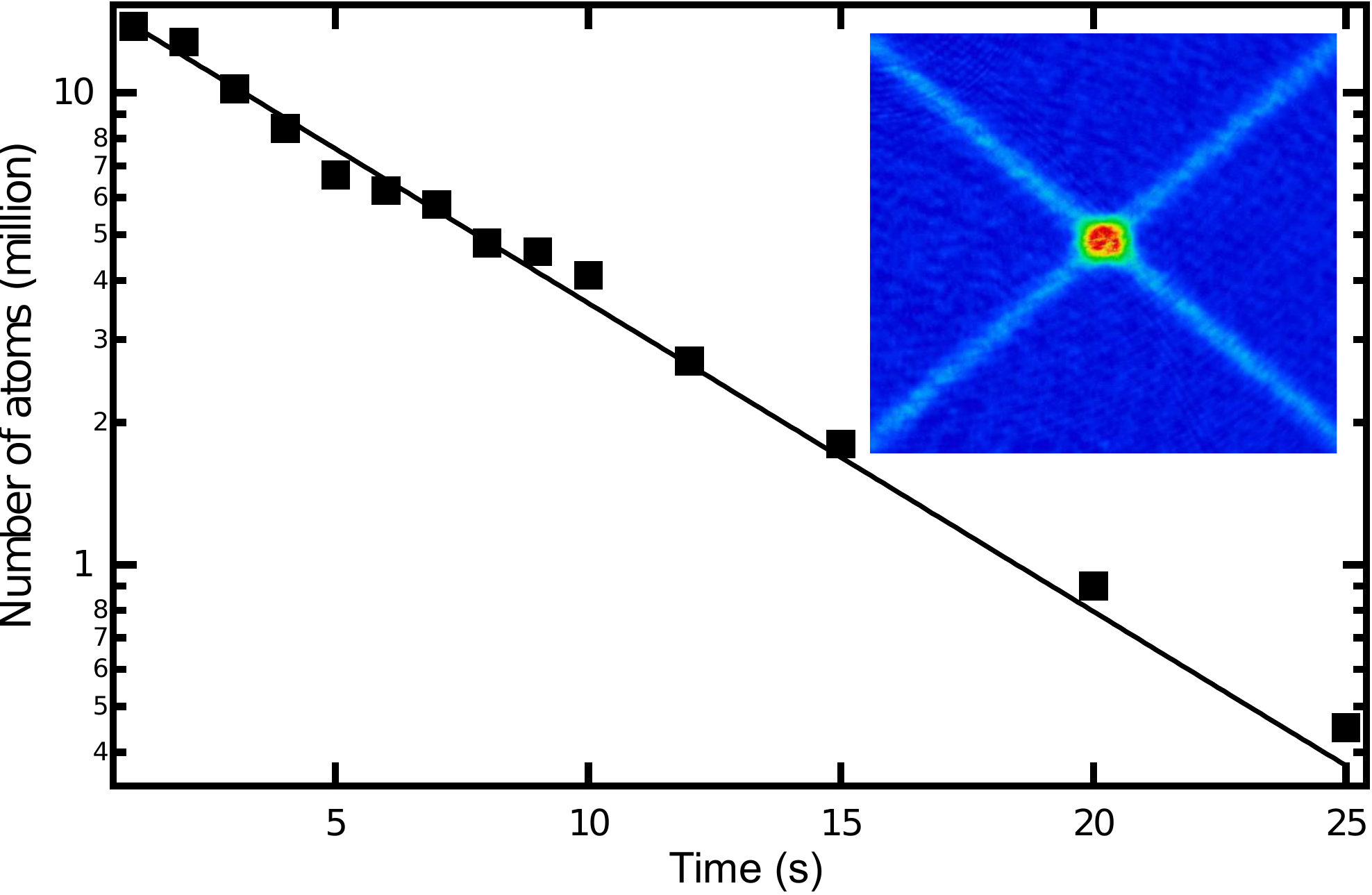}
\caption{Number of atoms in the trap versus trapping time. The solid line is an exponential fit for one body losses.  The inset shows an image of the trapped atomic sample after 500 ms.}
\label{fig:nb_at_vs_time}
\end{center}
\end{figure}

\section{Heterodyne non-demolition measurement}

\subsection{Collective measurement}

An ideal quantum non-demolition measurement can be seen as the textbook quantum projective measurement realized on an operator that commutes with the hamiltonian. In the case of multi-particles measurement, the result of the measurement \textit{collectively} constrains the sample and not individual atoms. Correlations between atoms have been created in what is called the measurement back-action. The back-action collapses the wavefunction and can be used to prepare spin-squeezed states \cite{Vanderbruggen2010}.

In the following, we consider an ensemble of $N_{\rm at}$ two levels atoms $\left\{ \left| a_{i} \right\rangle,\left| b_{i} \right\rangle \right\}$. We introduce the collective-spin operators:
\begin{eqnarray}
\widehat{J}_{x}& = & \frac{1}{2} \sum_{i} \left( \left| b_{i} \right\rangle \left\langle a_{i} \right| + \left| a_{i} \right\rangle \left\langle b_{i} \right| \right), \\
\widehat{J}_{y}& = & \frac{i}{2} \sum_{i} \left( \left| a_{i} \right\rangle \left\langle b_{i} \right| - \left| b_{i} \right\rangle \left\langle a_{i} \right| \right), \\
\widehat{J}_{z}& = & \frac{1}{2} \sum_{i} \left( \left| b_{i} \right\rangle \left\langle b_{i} \right| - \left| a_{i} \right\rangle \left\langle a_{i} \right| \right),
\end{eqnarray}
which obey the commutation relations $\left[ \widehat{J}_{i},\widehat{J}_{j} \right]=i \epsilon_{i j k} \widehat{J}_{k}$. The observable of the measurement is $\widehat{J}_{z} = \Delta \widehat{N}/2$, which is one half of the population difference between the two internal atomic states. For a coherent spin state (CSS) polarized along $\widehat{J}_{x}$, one has $( \langle \widehat{J}_{x} \rangle,  \langle \widehat{J}_{y} \rangle,  \langle \widehat{J}_{z} \rangle )=\left( N_{\rm at}/2,0,0\right)$, and $( \Delta {\widehat{J}_{x}}^{2} ,  \Delta {\widehat{J}_{y}}^{2}, \Delta {\widehat{J}_{z}}^{2} )=\left( 0,N_{\rm at},N_{\rm at} \right) / 4$.

A non-demolition measurement based on optical techniques requires to reach a high signal-to-noise detection together with a low light level to limit the decoherence induced by spontaneous emission. Spontaneous emission can be reduced using off-resonance probing, which is a dispersive measurement \cite{Kuzmich1998}. In that case, the off-resonance modification of the refractive index is measured by the phase shift induced on the probe optical field. Several techniques can be implemented to measure this phase shift, including  Mach-Zehnder interferometry \cite{Louchet2010}, mapping phase fluctuations into intensity fluctuations using a cavity tuned on the  side of its resonance \cite{Schleier2010}, or comparing the probe dephasing to a far from resonance local oscillator \cite{Teper08,Vanderbruggen2010}. The latter technique, called heterodyne detection, is adopted in the following.

\begin{figure}
\begin{center}
\includegraphics[width=13cm,keepaspectratio]{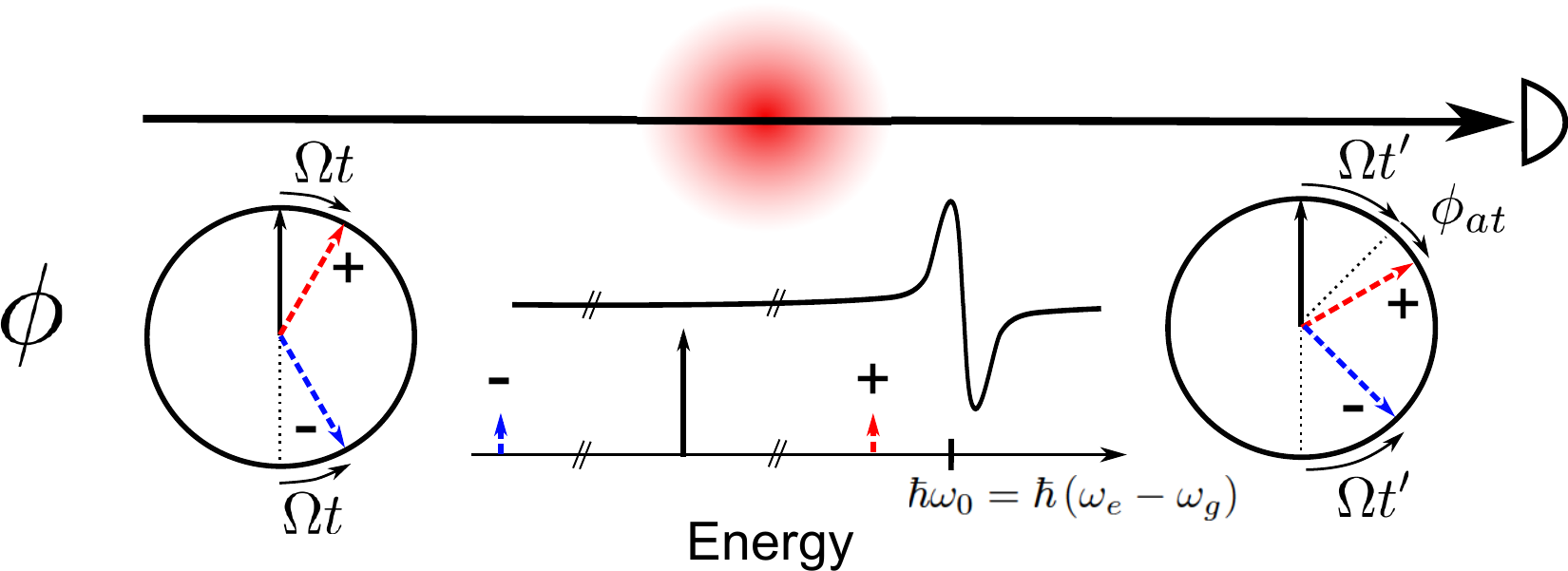}
\caption{Schematic view of a population sensitive heterodyne measurement. The circles are instantaneous phase representations of the optical frequency components in the frame rotating at the carrier frequency. The positive sideband (+, in red) starts in phase with carrier and turns with angular frequency $\Omega$ while the negative sideband (-, in blue) starts in opposite phase and turns with angular frequency $-\Omega$. The positive sideband is close to the atomic transition $\omega_0$, it accumulates an additional phase $\phi_{\rm at}$ that is extracted after demodulation.}
\label{fig:hetero_scheme}
\end{center}
\end{figure}

The detection scheme considered is based on frequency modulation spectroscopy \cite{drever83,Black2001,Bjorklund1983}. A laser beam is phase modulated to produce frequency sidebands; one sideband is placed close to an atomic transition and experiences a phase shift $\phi_{\rm at}$ passing through the atomic sample. The detection of the beat note at the modulation frequency then allows us to estimate the atomic population of the probed state. 

This scheme highly benefits from the almost noninteracting carrier used as a local oscillator. For a laser, limited by its quantum fluctuations (section \ref{sec:opt_det}), the signal-to-noise ratio (SNR) of the detection is
\begin{equation}
\textrm{SNR}\propto \frac{\sqrt{N_{s} N_{c}}}{\sqrt{(\sqrt{N_{s}})^2+(\sqrt{N_{c}})^2+(\sqrt{N_{e}})^2}} \sin{\phi_{\rm at}},
\label{eq:SNR}
\end{equation}
where $N_{s}$ represents the total number of photon detected in the sideband and $N_{\rm c}$ the photon number in the carrier. The contribution of technical noises in units of photon number is $N_{e}$. We underline that the two first terms of the denominator in equation (\ref{eq:SNR}) are the quantum phase noise of the laser expressed in terms of photon number. For a high power in the local oscillator, the shot noise overcomes the detection noise $N_{c} \gg  N_{e}$, and  equation (\ref{eq:SNR}) simplifies to $\mathrm{SNR} \approx \sqrt{N_{s}} \sin \phi_{\rm at}$, which corresponds to a sideband shot noise limited detection, independently of the sideband power.
The probe sensitivity to the fluctuations of the optical path is considered in the next section.

\subsection{A low sensitivity to common mode noise}
\label{sec:low_noise}

In homodyne detections, such as Mach-Zehnder interferometry, the sensitivity to path length fluctuations is determined by the optical wavelength $\lambda$. In a heterodyne detection scheme for which frequency components are spatially overlapped, the relevant length is the modulation wavelength $\lambda_{\rm mod}=\Omega/(2 \pi c)$. For a modulation in the microwave range, $\lambda_{\rm mod}$ is of the order of a few centimeters, whereas $\lambda$ is of the order of a micrometer. As a consequence, the sensitivity to length fluctuations is greatly reduced.

In the following, we focus on the sensitivity of a phase modulated detection with respect to optical path length fluctuations, and consider the effect of residual amplitude modulation (RAM) \cite{Gehrtz85}.
An optical field $E$, phase modulated at $\Omega$, can be written as:
\begin{equation} 
E=E_{0} \sum_{n=-\infty}^{\infty} J_{n}(\beta) \cos \left( \omega_{c}t+ n \Omega t \right),
\label{eq:Bessel}
\end{equation}
where $J_{n}$ are Bessel functions of the first kind and $\beta$ is the modulation depth.

For a small modulation depth ($\beta \ll 1$) the expansion can be limited to the first orders $\left( J_{0}(\beta), \; J_{1}(\beta), \; J_{-1}(\beta) \right) \approx (1,\; \beta, \; -\beta)$.	Let $\phi_{0}$, $\phi_{1}$, and $\phi_{-1}$ be the different atomic phase shift experienced by each frequency component of the optical field. The photocurrent $i_{\rm det}$ at the output of the photodetector is
\begin{equation}
	i_{\rm det}=\eta P_{\rm opt}\left(1+\beta\cos \left( \Omega t - \frac{\Omega L}{c} \right) \Delta \Phi_{+} + \beta \sin \left( \Omega t - \frac{\Omega L}{c}\right) \Delta \Phi_{-}\right),
\end{equation}
where $L$ is the distance between the modulator and the detector, $P_{\rm opt}$ the optical power in the local oscillator, $\eta$ the detection sensitivity of the photodiode,  and
\begin{eqnarray}
	\Delta \Phi_{+} & = & \cos \left( \phi_{1} - \phi_{0} \right) - \cos \left( \phi_{0} - \phi_{-1} \right), \label{eq:phi+} \\
	\Delta \Phi_{-} & = & \sin \left( \phi_{1} - \phi_{0} \right) - \sin \left( \phi_{0} - \phi_{-1} \right). \label{eq:phi-}
\end{eqnarray}
After the demodulation of $i_{\rm det}$ by $\sin(\Omega t + \Phi_{\rm dem})$ with $\Phi_{\rm dem}=\Omega L/c$, and for length noise fluctuations $\delta L$ of the optical path integrated in the detection bandwidth $\Delta f$, the dispersive signal obtained is $S= S_0+\delta S_L$, where:
\begin{eqnarray}
	S_0= \eta \beta P_{\rm opt} \Delta \Phi_{-},\\
	\delta S_L = 2 \pi \eta \beta P_{\rm opt} \frac{\delta L}{\lambda_{\rm mod}} \Delta \Phi_{+}.
	\label{eq:sig_demod}
\end{eqnarray}
$\delta S_L$ is the component of the noise that is due to fluctuations of the optical path length.\\
Since the atomic contributions to the phase shift are small $\phi_{-1}$, $\phi_{0}$, $\phi_{1} \ll 1$, we have at second order:
\begin{eqnarray}
	\Delta \Phi_{+} & = & \frac{1}{2} \left( \phi_{1} - \phi_{-1} \right) \left( 2 \phi_{0} - \phi_{1} - \phi_{-1} \right),\label{eq:atomic_phase1} \\
	\Delta \Phi_{-} & = & \phi_{1} + \phi_{-1} - 2 \phi_{0}.
	\label{eq:atomic_phase2}
\end{eqnarray}
Considering the case depicted in figure \ref{fig:hetero_scheme}, we have $\phi_{1}=\phi_{\rm at}\ll 1$ and $\phi_{-1}=\phi_{0}=0$. 
From equations~(\ref{eq:sig_demod}), (\ref{eq:atomic_phase1}) and (\ref{eq:atomic_phase2}), the detection noise $\delta S_L$ is $\eta \beta P_{\rm opt}\phi_{\rm at}^2\delta L/\lambda_{\rm mod}$. Hence, for a pure phase modulation, a small atomic phase shift $\phi_{\rm at}$ makes the signal even less sensitive to length fluctuations than $\delta L/\lambda_{\rm mod}$.\\

A major noise source in this heterodyne scheme arises from the residual amplitude modulation  due to phase modulation imperfections. The residual amplitude modulation unbalances the field amplitude in the sidebands and the optical electric field can be written as $E=E_0 e^{i \omega_c t}\left (1+\beta(1+\epsilon) e^{i \Omega t}-\beta(1-\epsilon) e^{-i \Omega t}\right)$. The signal after demodulation is:
\begin{equation}
S= \eta \beta P_{\rm opt} \left[(\Delta \Phi_{-}+\epsilon \Delta \Phi_{-,\rm AM}) + 2 \pi \frac{ \delta L}{\lambda_{\rm mod}}(\Delta \Phi_{+}+\epsilon \Delta \Phi_{+,\rm AM})\right],
\end{equation}
where $\Delta \Phi_{+,\rm AM}$ and $\Delta \Phi_{-,\rm AM}$ are the atomic phase contributions for a pure amplitude modulation:
\begin{eqnarray}
		\Delta \Phi_{+,\rm AM} & = & 2 + \frac{1}{2} \left[ \left( \phi_{1} - \phi_{0} \right)^{2} + \left( \phi_{-1} - \phi_{0} \right)^{2} \right], \\
		\Delta \Phi_{-,\rm AM} & = & \phi_{1} - \phi_{-1}.	
\end{eqnarray}
In the case of figure \ref{fig:hetero_scheme}, the overall noise contribution of optical path  length fluctuations is:
\begin{equation}
\delta S_L=\eta \beta P_{\rm opt}(\phi_{\rm at}^2+\epsilon)\frac{\delta L}{\lambda_{\rm mod}}.\label{eq:noise_length} 
\end{equation}

%The total noise variance $\delta S^2$ of the detection in a bandwidth $\Delta f$ is the contribution of the optical shot noise $\delta S_{\rm shot}$, the electronic noise $\delta S_e$, and the path length fluctuation noise $\delta S_L$:
%\begin{equation}
%\delta S^2= \delta S_{\rm shot}^2+\delta S_e^2 + \delta S_L^2= 2 e \eta P_{\rm opt} \Delta f+\delta S_e^2 + \left(\eta \beta P_{\rm opt}(\phi_{\rm at}^2+\epsilon)\right)^2\left(\frac{\delta L}{\lambda_{\rm mod}}\right)^2.
%\end{equation}\\
%To realize an optimal detection, the essential requirement is to overcome the electronic noise with the optical shot noise, $P_{\rm opt} \gg \delta S_e^2/(2 e \eta \Delta f)$.
As a consequence, in a phase modulated detection, the length fluctuations are rejected by a factor $(\phi_{at}^2+\epsilon)\lambda/\lambda_{\rm mod}$ with respect to a Mach-Zehnder interferometer (equation (\ref{eq:noise_length})). For our experimental parameters, $\epsilon=10^{-2}$, $\lambda_{\rm mod}$=10 cm, $\lambda=1~\upmu$m, and for $\phi_{at}<100$ mrad, it represents 7 orders of magnitude of rejection on $\delta S_L$.
In addition, $\delta S_L/\delta S_{\rm shot}$ scales as $\sqrt{P_{\rm opt}}$. As a result, if the photodetector is shot noise limited for a low optical power in the local oscillator (section \ref{sec:opt_det}), there is no need of a dynamical control of the optical path length.

\subsection{Optical set-up}

\begin{figure}
\begin{center}
\includegraphics[width=12cm,keepaspectratio]{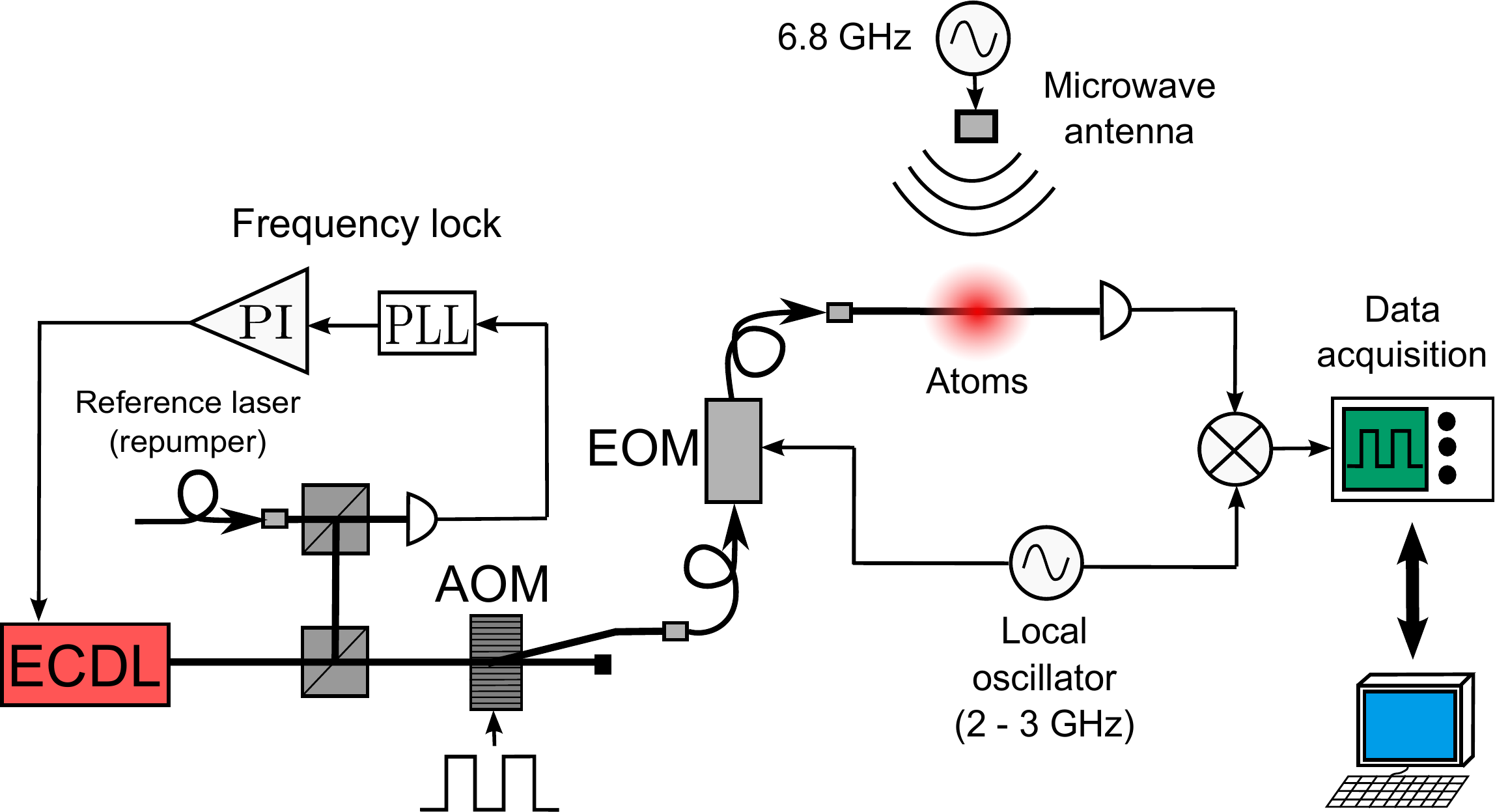}
\caption{Set-up of the single-pass non-destructive measurement. The probe light is generated by an extended cavity diode laser and offset locked to the atomic transition via a phase locked loop (PLL) and a proportional integrator feedback loop. The absolute frequency of the light is controlled by changing the reference frequency of the PLL. The atoms are probed with the optical pulses generated by the AOM. The sidebands for the heterodyne detection are inserted by a phase modulator.}
\label{fig:QND_bench}
\end{center}
\end{figure}

The optical set-up is presented in figure \ref{fig:QND_bench}. The detection beam is generated by an extended cavity diode laser, frequency locked to the reference laser \cite{appel09}. The detuning of the carrier, and thus of the sideband from the atomic resonance, is electronically controlled by a Phase Locked Loop (PLL).
An acousto-optic modulator (AOM) is used as a switch for the detection beam and generates pulses with $300$ ns rise time. A polarization maintaining fibered electro-optic modulator (EOM) produces frequency sidebands that probe the atomic sample. The beating signal of the carrier with the sidebands is detected on a fast photodiode (see section.~\ref{sec:opt_det}), and the demodulated signal is acquired with a digital oscilloscope. 
The internal atomic state is manipulated with a microwave signal at about $6.834$ GHz generated by a home-made frequency chain. This chain relies on a low noise quartz oscillator (Wenzel Associates, mod. Blue Top Ultra Low Noise Oscillator) which has -113dBc/Hz phase noise at 1 Hz. After a suitable amplification, the microwave is shined on the atoms with an antenna placed at about 20 cm from the sample.
\subsection{The optical detector}
\label{sec:opt_det}

\begin{figure}
\begin{center}
\includegraphics[width=10cm,keepaspectratio]{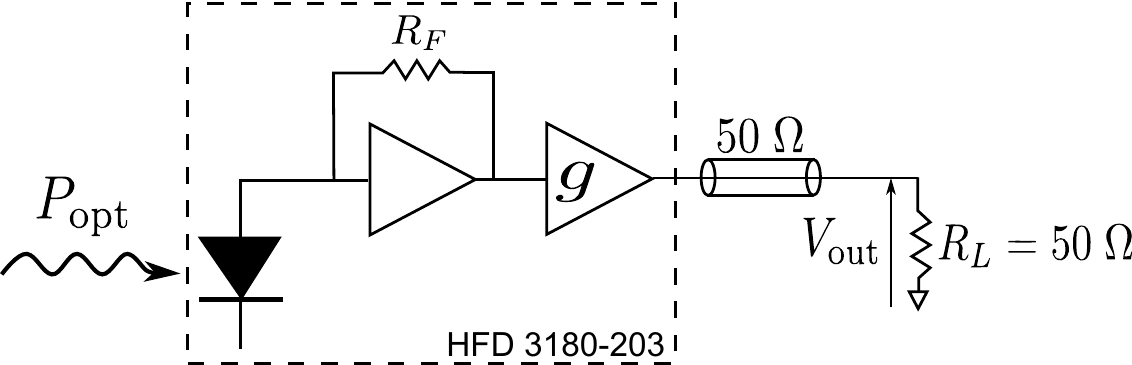}
\caption{Scheme of the optical detector. It is composed of a fast photodiode followed by a transimpedance amplifier of gain $R_{F}$ and a buffer with gain $g$.}
\label{fig:photodiode}
\end{center}
\end{figure}

The beat-note signal is detected on a fast GaAs PIN photodiode with integrated transimpedance amplifier designed for high-speed optical communications at 850 nm (Finisar, model. HFD3180-203). The electronic scheme for the photodiode and the amplifier is presented in figure \ref{fig:photodiode}. The integration of the photodiode and amplifier in the same package greatly reduces parasitic capacitances and allows a high detection bandwidth of a few GHz together with a high transimpedance gain of the order of one k$\Omega$. The detector is mounted on a home-made high frequency printed circuit board. The output of the amplifier is connected to a 50 $\Omega$ matched coplanar waveguide and AC-coupled to a SMA connector.
\begin{figure}
\begin{center}
\includegraphics[width=12cm,keepaspectratio]{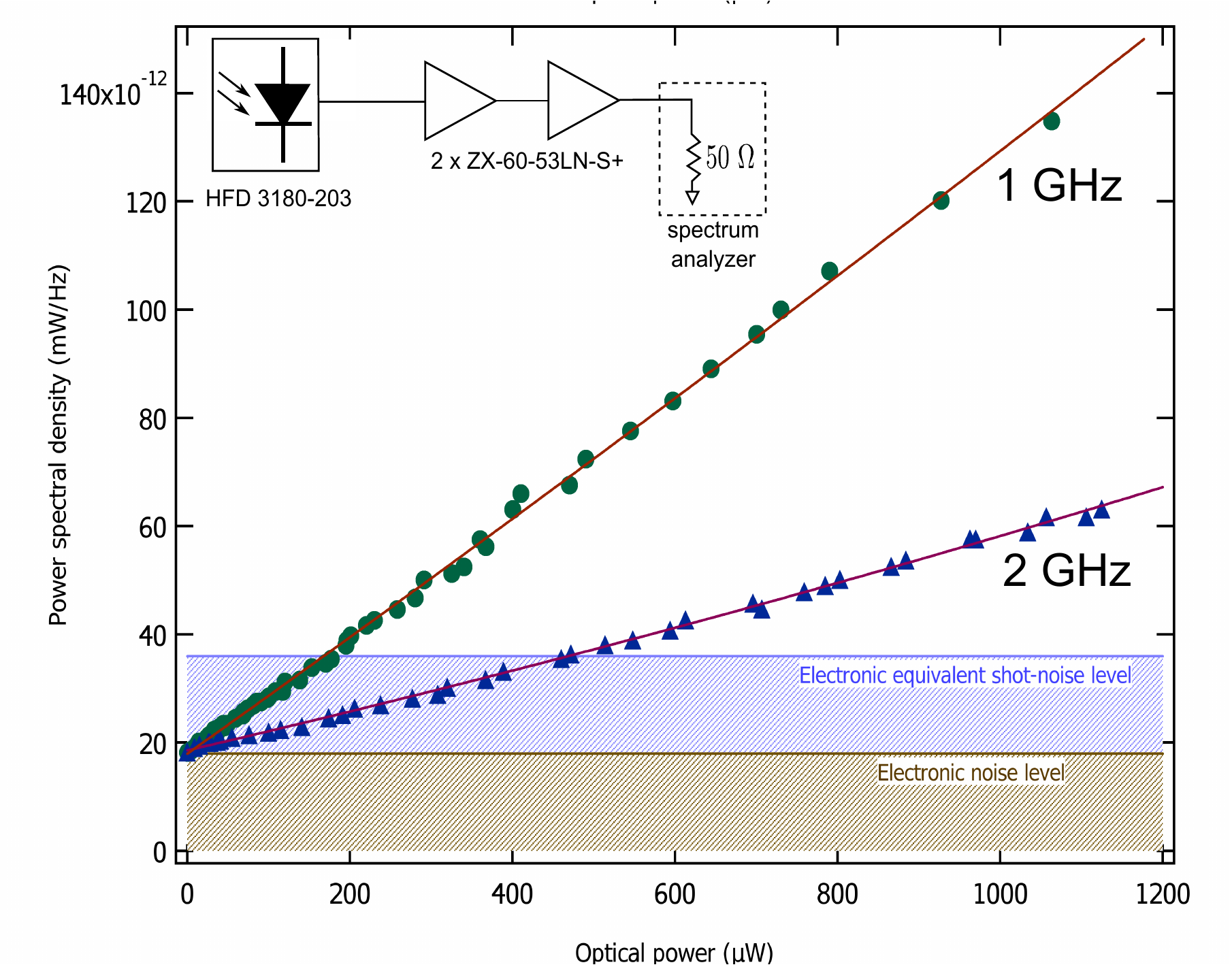}
\caption{Noise power spectral density versus the incident optical power for two different detection frequencies, 1 GHz (circles) and 2 GHz (triangles). The inset presents the set-up used for the noise measurement.}
\label{fig:noise_pd}
\end{center}
\end{figure}

The noise at the output of the photodiode has been measured with a spectrum analyzer. The output of the transimpedance amplifier was amplified by $36$ dB to overcome the noise level of the spectrum analyzer. The noise power spectral density  of the illuminated detector was measured at 1 GHz and 2 GHz, as reported in figure \ref{fig:noise_pd}.

The measurement shows a very good linearity of the noise power spectral density versus the incident power, which means that both the laser and the detector are shot-noise limited at the adopted frequencies. The figure of merit $\kappa$ considered  is the electronic noise equivalent light shot-noise, that is the optical power required to generate the same noise as the detection electronics \cite{windpassinger09}. We measured $\kappa=165~\upmu$W at 1 GHz, and $\kappa=469~\upmu$W at 2 GHz. 

Moreover, the slope provides a direct measurement of the photodiode gain $G_{\rm PD}=V_{\rm out}/ P_{\rm opt}= g R_{F} \eta$ where $\eta \sim 0.5$ A/W is the sensitivity of the photodiode (see figure \ref{fig:photodiode} for notations). At the output of the transimpedance amplifier, voltage fluctuations induced by the light shot noise are $v_{n} = g R_{F} \sqrt{2 e \eta P_{\rm opt} \Delta f}$, where $e$ is the charge of the electron and $\Delta f$ the detection bandwidth. Hence the noise power spectral density on a resistance load $R_{L}$ of $50 \; \Omega$ is:
\begin{equation}
\mathrm{PSD}_{n} = \frac{v_{n}^2}{R_{L}\Delta f}=2 e \frac{G_{\rm PD}^{2}}{R_{L} \eta} P_{\rm opt},
\end{equation}
From the slopes of figure \ref{fig:noise_pd}, one gets $G_{\rm PD} = 1466$ V/W at 1 GHz and $G_{\rm PD} = 880$ V/W at 2 GHz, values in good agreement with the nominal value of $1250$ V/W.

\section{Non-demolition measurement of an evolving atomic state.}

We now switch to the application of our heterodyne scheme to the measurement of a cold atomic cloud in free fall during the measurement.
The spontaneous emission rate induced by the probe is first measured. Then, a real-time measurement of atoms undergoing Rabi oscillations is realized. Finally, the detection tool is used to follow continuously the evolution of an atomic state during an interferometric sequence.

\subsection{Effective scattering rate}

\begin{figure}
\includegraphics[width=1\textwidth]{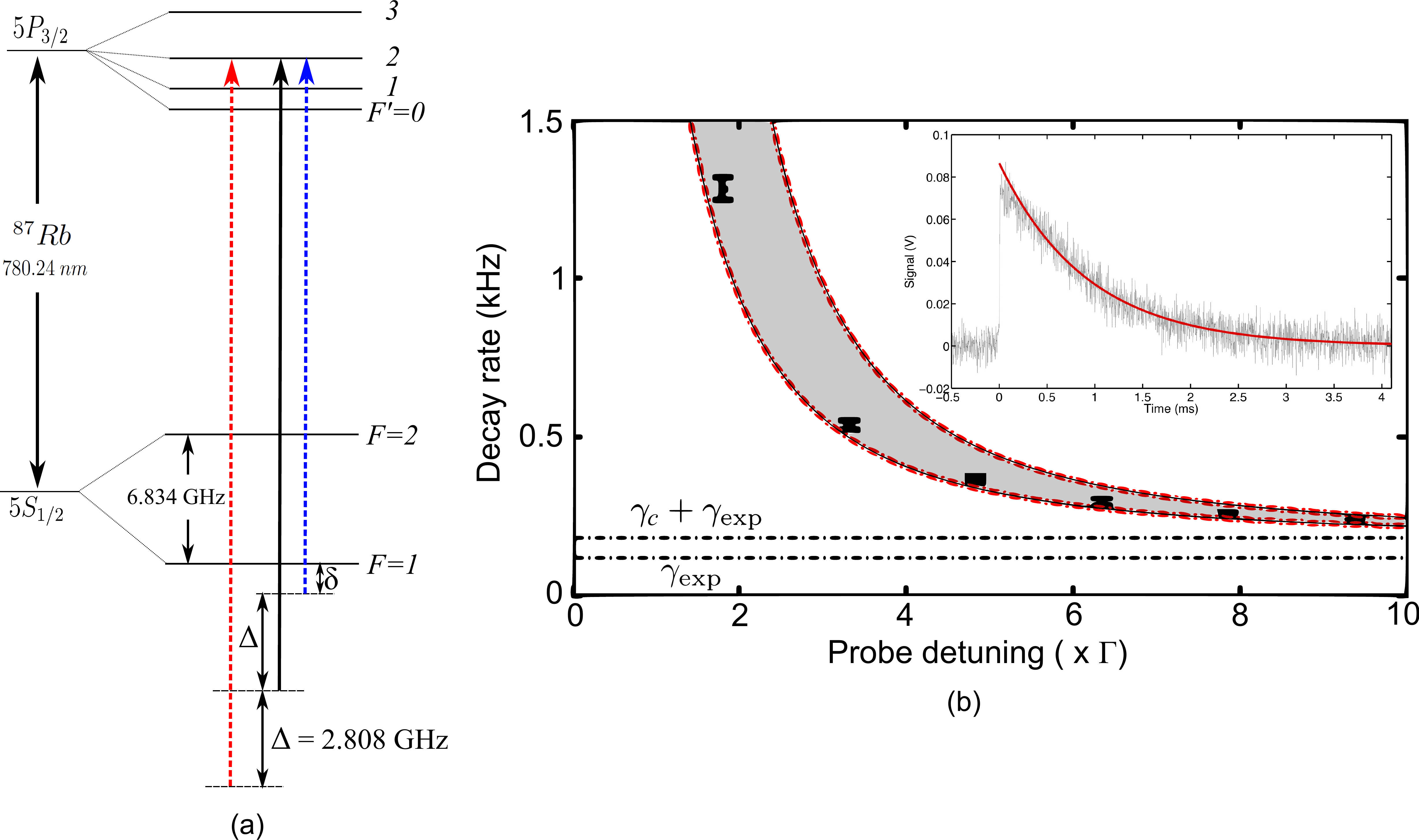}
\caption{Destructivity as a function of probe detuning. (a) Frequency position of the optical triplet with respect to the $D_2$ atomic line. (b) Probe induced scattering rate. The gray shaded area, with red dashed borders represents the theoretical decay rate. The only adjustable parameter is the expansion rate of the cloud ($\gamma_{\rm expansion}$). In inset is shown the demodulated atomic signal at $\delta=4.81\Gamma$, with the relative exponential fit.}
\label{fig:spontaneous}
\end{figure}

In experiments that concern measurement induced squeezing, the probe induced spontaneous emission is among the limiting factors to reach highly entangled states \cite{Echaniz2005}. To quantify the destructivity induced by the probe in our set-up, the atoms prepared in the $\left|F=1\right\rangle$ hyperfine state are released from the dipole trap and continuously probed. The cloud contains 4.7(3) million atoms in an initial rms size of $\sigma_{\rm at}=41~\upmu$m, and expands with a temperature of $55(5)~\upmu$K. The optical frequency detunings with respect to the atomic transitions are presented in figure \ref{fig:spontaneous}(a). The modulation frequency is fixed to 2.808 GHz. The detuning ($\delta$) of the probing sideband with respect to the atomic transition is changed by moving the absolute frequency of the carrier. As a consequence, the quadrature set by the demodulation phase is independent of the detuning $\delta$.\\

Spontaneous emission induced by the probe optically pumps the atoms to $\left|F=2\right\rangle$ where they become transparent for the probe. The measured exponential decay time $\tau$ of the signal (inset of figure \ref{fig:spontaneous}(b)) expressed as a decay rate $\gamma=1/(\pi\tau)$ is then closely linked to the spontaneous emission rate of the probe.

The predicted decay rate $\gamma$, shown in figure \ref{fig:spontaneous}(b) as a gray zone, is calculated from the beam waist ($245~\upmu$m), the carrier power ($P_c=120~\upmu$W), and  the power in each sideband ($P_s=76~$nW) of the linearly polarized probe, and is expressed as:
\begin{equation}
	\gamma=\gamma_{s}+\gamma_{c}+\gamma_{\rm exp}=b_2\frac{\Gamma I_{s}/(2 I_{\rm sat,2})}{1+4\left( \frac{\delta}{\Gamma} \right)^2+I_{s}/I_{\rm sat,2}}+\sum_{i=0}^{2} b_i \frac{\Gamma I_{c}/(2 I_{\rm sat,i})}{1+4\left( \frac{\Delta_i}{\Gamma} \right)^2+I_{c}/I_{\rm sat,i}}+\gamma_{\rm exp}\label{eq:spont},
\end{equation}
where $I_{\rm sat,i}$ is the saturation intensity for the $\pi$ transition from $\left| F=1 \right\rangle$  to $\left| F'=i \right\rangle$, and the Zeeman sub-levels are considered equally populated. We have $I_{\rm sat,0}=16.67~\textrm{W}/\textrm{m}^2$, $I_{\rm sat,1}=26.7~\textrm{W}/\textrm{m}^2$, and $I_{\rm sat,2}=61.23~\textrm{W}/\textrm{m}^2$. The branching probabilities $b_i$ to spontaneously scatter from $\left|F'=i \right\rangle$ to $\left|F=2 \right\rangle$ are $b_0=0$, $b_1=1/5$, and $b_2=1/2$. $I_{s}=2P_{s}/(\pi w^2)$  is the sideband intensity while $I_{c}$ is the carrier intensity, $\Delta_i=2.808+\Delta_{2i}~$GHz is the carrier detuning from the transition and $\Delta_{2i}$ is the angular frequency difference between $\left|F'=2\right\rangle$ and $\left|F'=i\right\rangle$. $\gamma_{\rm exp}=120~$Hz is a constant to take into account the signal loss due to the combination of fall and expansion of the cloud; it is the only adjusted parameter in figure \ref{fig:spontaneous}(b). A simulation of the loss of signal with the expansion gives a decay rate of $159~$Hz, showing good qualitative agreement with the experimental value.

The gray zone in figure \ref{fig:spontaneous}(b) represents the expected region for the decay rate versus the probe detuning when uncertainties in the experimental parameters are considered. It shows a good agreement with the experimental data. 
Although the probe is very weak compared to the carrier, it dominates the spontaneous emission rate for $\delta<5\Gamma$ in this specific experimental realization (figure \ref{fig:spontaneous}(b)).

\subsection{Rabi oscillations}
\label{sec:Rabi}
\begin{figure}
  \centering
  \includegraphics[width=1\textwidth]{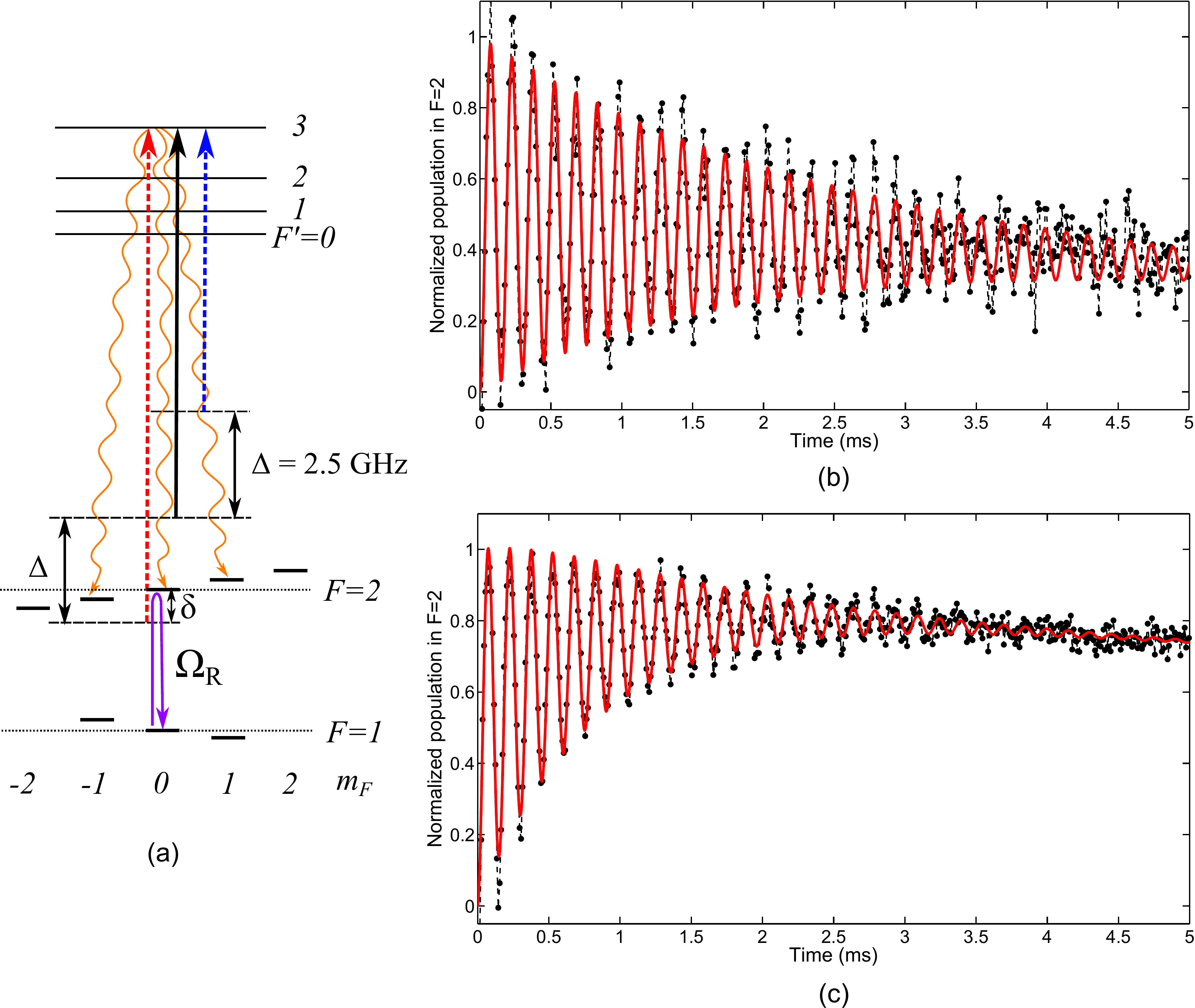}
  \caption{Non-demolition measurement of the atomic population in $\left|F=2\right\rangle$ when Rabi oscillations are driven in the atomic sample. The probing sequence is realized with a $1.25~\upmu$s long pulse repeated every $10~\upmu$s. Each graph is realized in a single experimental cycle. The detuning of the probe to the transition is set to $7.9 \Gamma$ (a) and $0.8 \Gamma$ (b). The red curve is a fit to the data and the dashed black line helps as a guide to the eyes.}
  \label{fig:Rabi}
\end{figure}
Microwave driven Rabi oscillation is a well understood physical phenomenon and a typical demonstration of coherent manipulation. It is therefore very convenient to precisely characterize how a non-destructive probe affects the oscillating system \cite{Windpassinger2008,Souma2006,Smith2004}.

The atomic system is released from the MOT and optically pumped in the hyperfine ground state $\left| F=1,m_F=0 \right\rangle$. At $t=0$, a microwave driven Rabi oscillation is induced between $\left|F=1,m_F=0\right\rangle$ and $\left|F=2,m_F=0\right\rangle$ levels. A magnetic bias field of $0.5$ G is applied along the vertical axis so that other Zeeman transitions are off-resonance. The cloud contains $\sim10^7$ atoms and expands with a temperature of $80~\upmu$K. The probe used in this experiment has a waist of $w_{0}=800~\upmu$m and delivers pulses of $1.25~\upmu$s with a repetition rate of 100 kHz or 50 kHz. The carrier power is $70~\upmu$W and the sideband power is $90~$nW. As described in figure \ref{fig:Rabi}(a), one sideband is blue detuned by $\delta$ from the $\left|F=2\right\rangle \rightarrow \left|F'=3\right\rangle$ transition whereas the carrier is 2.5 GHz away from the resonance. Hence, the configuration is probing the population in the $\left|F=2\right\rangle$ state.\\

Figure \ref{fig:Rabi} shows two examples of real time measurements of the atomic evolution obtained in a single shot.
Figure \ref{fig:Rabi}(b) was taken with a probe detuning of $7.9\ \Gamma$ and shows a trace of Rabi oscillation undergoing decoherence. Figure \ref{fig:Rabi}(c), which was taken with a detuning of $0.8\ \Gamma$, presents an up-lift of the average oscillation. This effect is qualitatively explained by taking into account the probe induced spontaneous emission which transfers atoms from $\left| F=2,m_{\rm F}=0\right\rangle$ to $\left| F=2,m_{\rm F}\neq0\right\rangle$. There, the atoms still contribute to the detected signal while being off-resonance for the microwave excitation.\\

\begin{figure}
  \centering
  \subfloat[][]{\includegraphics[width=0.6\textwidth]{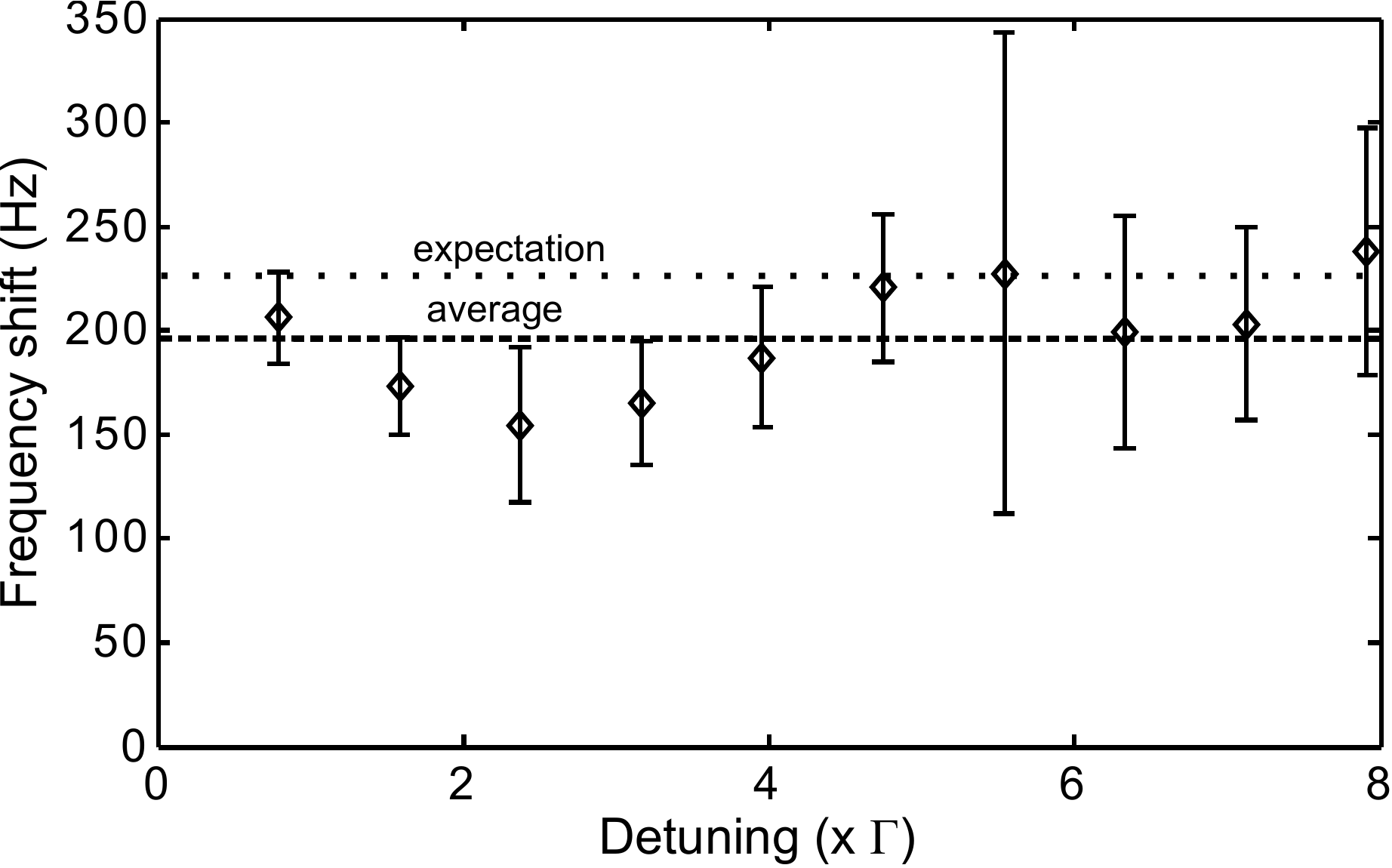}}\\
 \subfloat[][]{\includegraphics[width=0.61\textwidth]{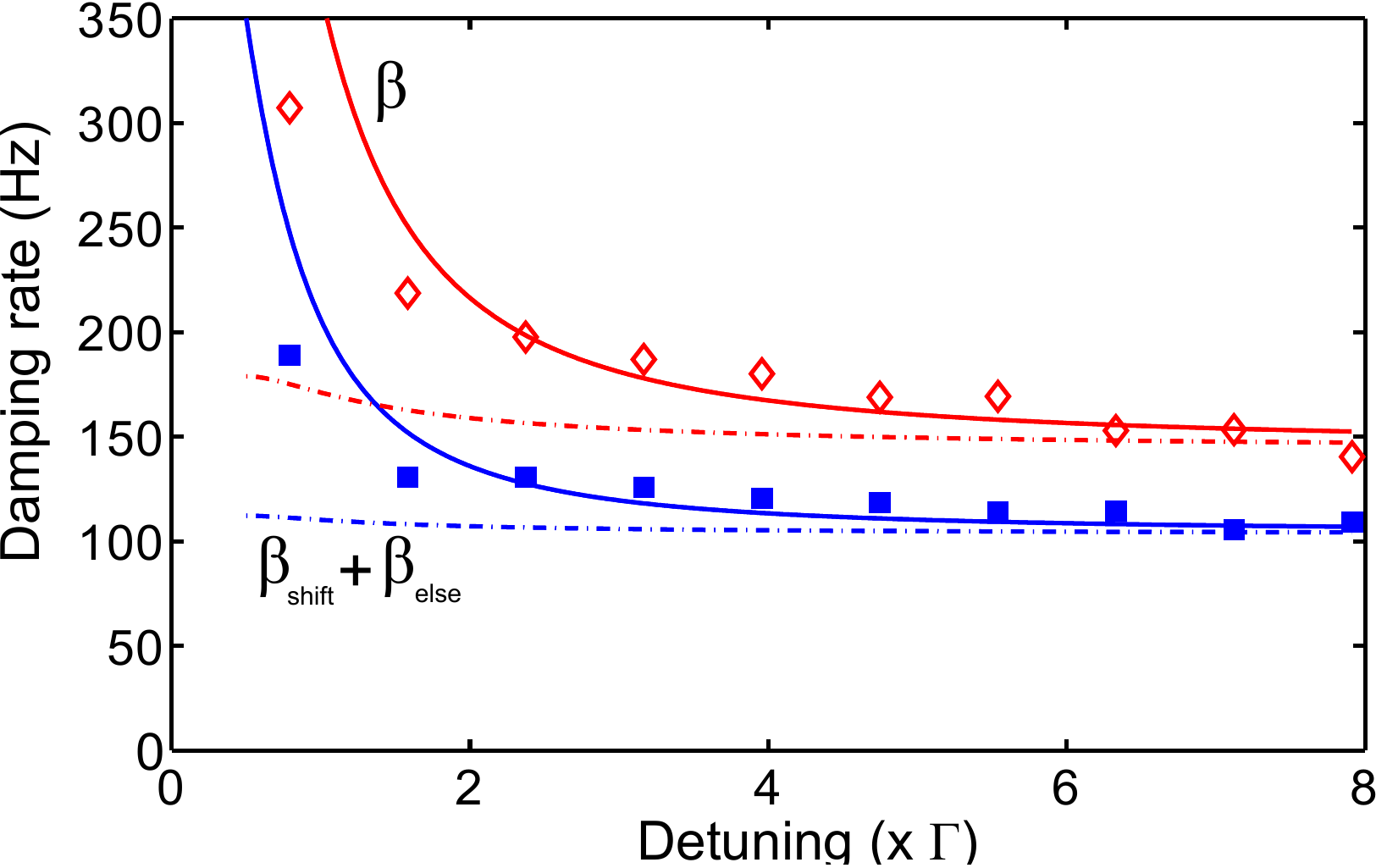}}
  \caption{Dependence of the Rabi oscillation parameters from the probe detuning. (a) Difference of the Rabi frequency for data sets having repetition rates of $100$ kHz and $50$ kHz. (b) Damping rate of the oscillations. In red open diamond are presented data with a repetition rate of  $100$ kHz, whereas blue squares represent data at $50$ kHz. The solid lines are the expected decay rates taking into account spontaneous emission, inhomogeneous light-shift, and expansion. The dashed lines are the decay rates without the spontaneous emission contribution.}
  \label{fig:Rabi_para}
  \end{figure}

Figure \ref{fig:Rabi_para}(a) presents the difference of Rabi oscillation frequency $\delta \Omega_{\rm R}$  for two different repetition rates of $100$ kHz and $50$ kHz as a function of the detuning ($\delta \Omega_{\rm R}=\Omega_{\rm R,100 \rm kHz}-\Omega_{\rm R,50 \rm kHz}$,). The Rabi frequency is obtained by fitting the first $800~\upmu$s of the oscillation where the expansion has little effect. It clearly shows that the probe light modifies the Rabi frequency around its unperturbed value of ($\Omega_{\rm R}\sim 6.6$ kHz). For the experimental parameters given above, the carrier induces a light shift $\Delta E_{\rm c}/h\sim2$ kHz for a $100$ kHz repetition rate whereas the sideband light-shift is below 380 Hz for the smallest detuning considered. The difference in Rabi frequency calculated from the carrier light shift is $\delta \Omega_{\rm R}=227$ Hz, in reasonable agreement with the $197$ Hz average of figure \ref{fig:Rabi_para}(a).
In addition, from the fitting procedure, we extract that a SNR=1 is obtained for a destructivity of $2.6\times 10^{-6}$ scattering event per atoms.

The exponential damping of the Rabi oscillations is described by its damping rate $\beta$ which is the sum of three independent terms:
\begin{equation}
\beta=\beta_{\rm spont}+\beta_{\rm shift}+\beta_{\rm else}.
\end{equation}
$\beta_{\rm spont}$ represents the damping of the oscillation by the spontaneous emission process of both the carrier and the sideband, $\beta_{\rm shift}$ results from the light shift inhomogeneity induced by the Gaussian light  profile \cite{Windpassinger2008}, which is essentially limited to the carrier effect, and $\beta_{\rm else}$ represents the probe independent decoherence effects such as the microwave inhomogeneity and the expansion of the cloud expansion. The behavior of $\beta$ versus the probe detuning is presented in figure \ref{fig:Rabi_para}(b).  The spatial inhomogeneity of the carrier light-shift contributes to damp the oscillation in a characteristic time that is proportional to the inverse of the Rabi frequency dispersion $\Delta \Omega_{\rm R}=\sqrt{\left\langle \Omega_{\rm R}^2\right \rangle-\left\langle \Omega_{\rm R}\right \rangle^2}$, where the averages are taken over the cloud profile. It results in a damping rate $\beta_{\rm shift}=\alpha\Delta E_{\rm c}^2/ \left(2 \hbar^2 \Omega_{\rm R} \right)$ where $\alpha$ characterizes the inhomogeneity over the cloud size, and depends on the atomic and probe sizes. The theoretical curves plotted in figure \ref{fig:Rabi_para}(b) are given for $\beta_{\rm else}=90~$Hz and $\alpha=0.162$.

The sum of the three damping coefficients reproduces well the observed behavior. Close to resonance, the damping of the oscillation is dominated by spontaneous emission events. Tuning the probe further from resonance greatly reduces the effect of spontaneous emission from the sideband, and the damping becomes limited by the inhomogeneous light shift induced by the carrier. 

\subsection{Interferometric phase in a spin-echo sequence}

We present here the non-demolition measurement of the atomic state during an interferometric sequence. We are particularly interested in the evolution of the atomic pseudo-spin in the equatorial plane of the Bloch representation, since it is directly associated to the evolution of the interferometric phase.

\begin{figure}
  \centering
    \subfloat[][]{
  \begin{minipage}[c]{0.47\textwidth}
  \centering
    \label{fig:interfero_data}\includegraphics[width=1\textwidth]{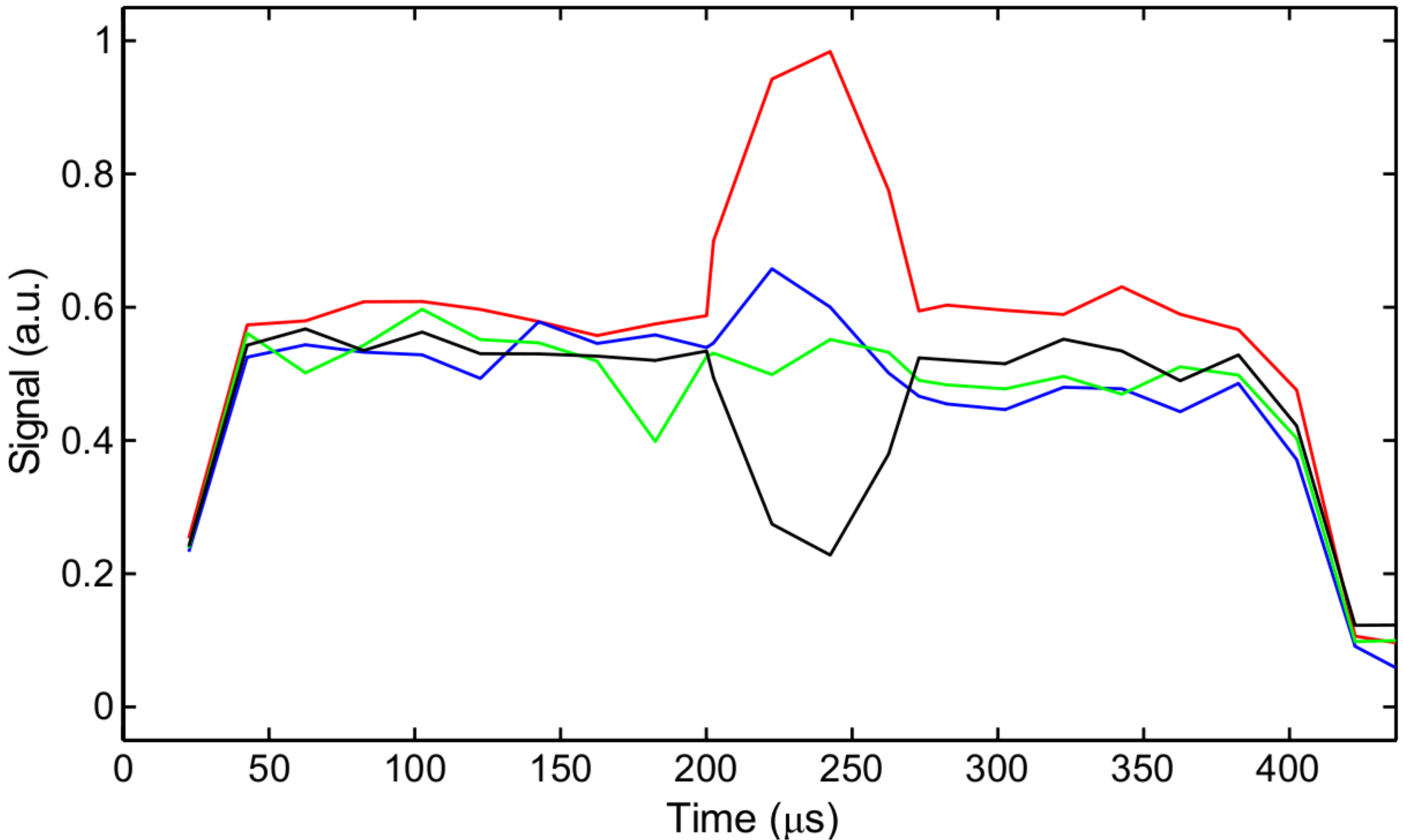}
    \end{minipage}}
    \hspace{0.3cm}
  \subfloat[][]{
  \begin{minipage}[c]{0.47\textwidth}
  \centering
    \label{fig:interfero_simul}\includegraphics[width=1\textwidth]{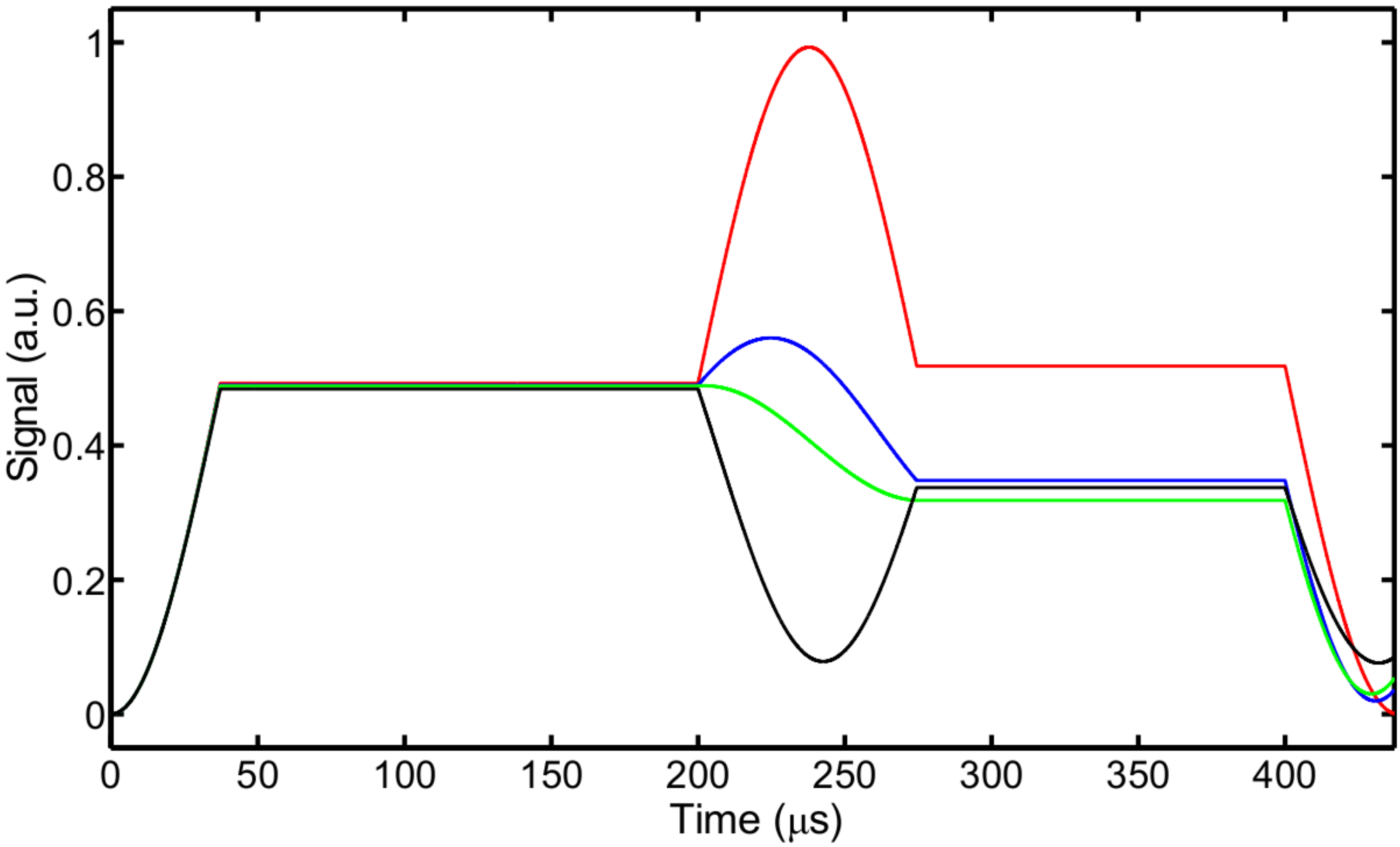}
  \end{minipage}}\\
  \vspace{0.1cm}
   \subfloat[][]{
  \begin{minipage}[c]{0.34\textwidth}
  \centering
    \label{fig:Blochsphere}\includegraphics[width=1\textwidth]{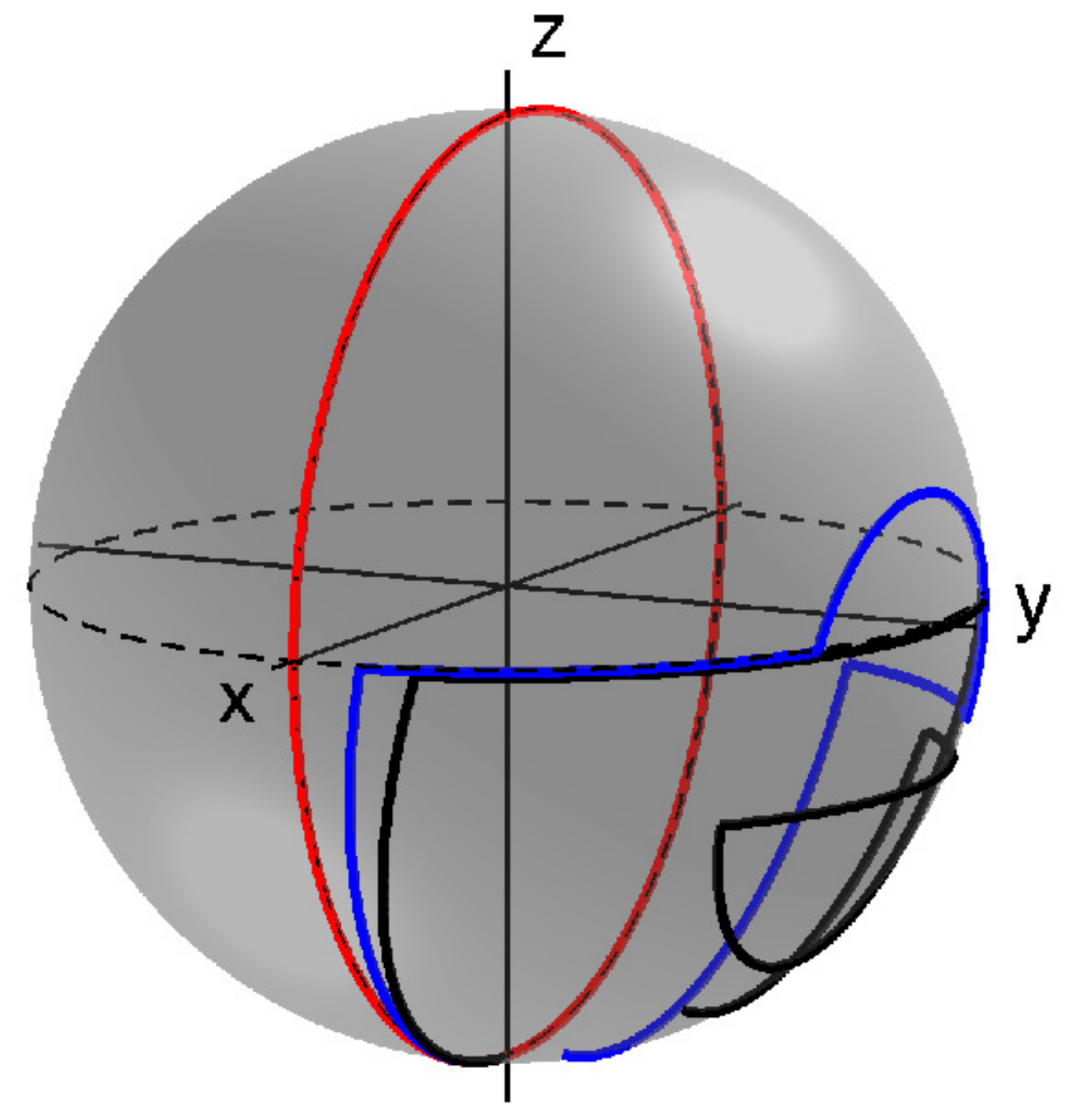}
    \end{minipage}}
\hspace{0.3cm}
  \subfloat[][]{
  \begin{minipage}[c]{0.60\textwidth}
  \centering
  \label{fig:interfero_amp}\includegraphics[width=1\textwidth]{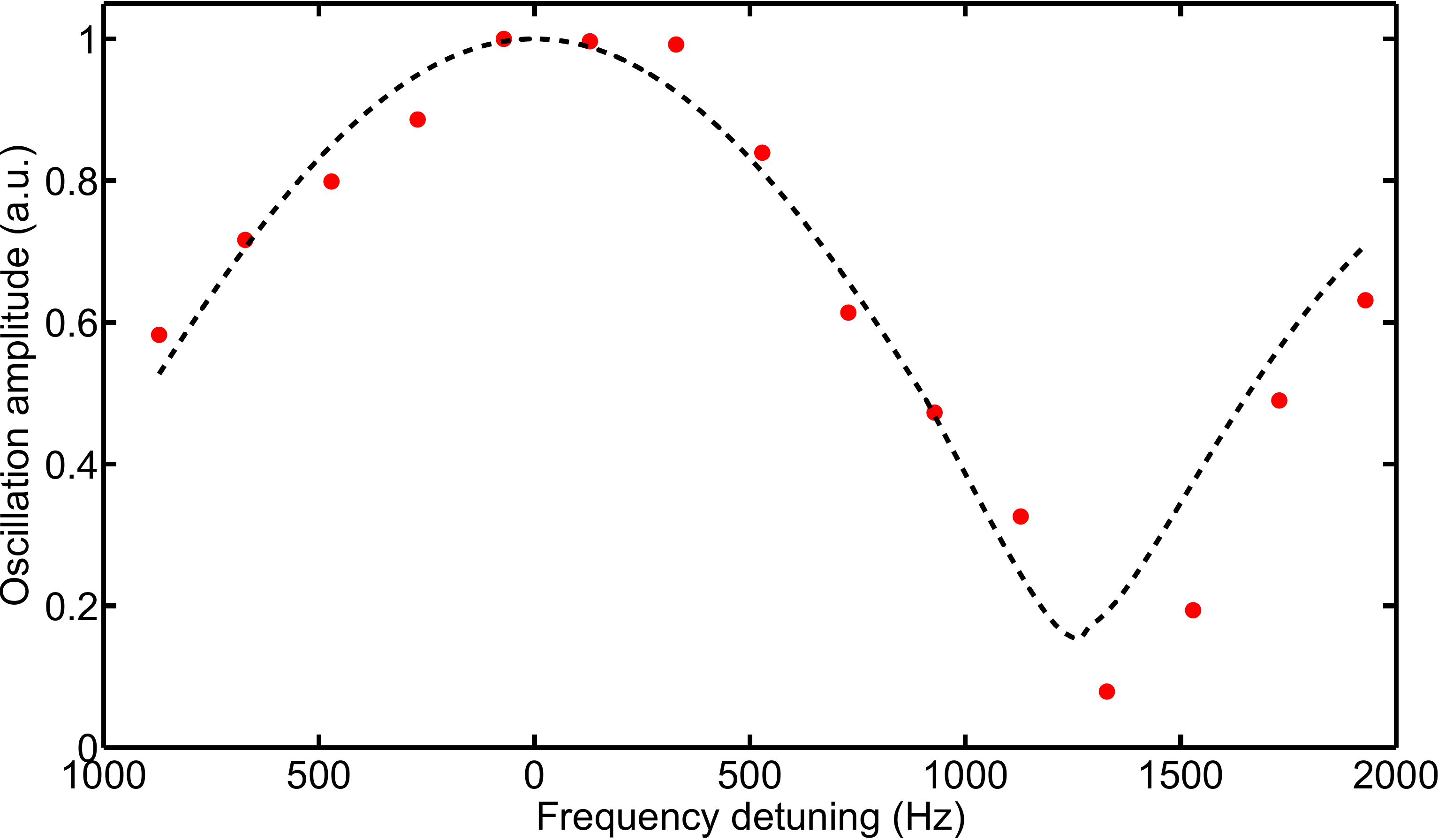}
  \end{minipage}}
  \caption{Spin-echo sequence. (a) Non-demolition measurement of the number of atoms in $\left| F=2\right \rangle$ during the interferometric sequence. Each curve is the average of three repetitions. The measurement uses $1.25~\upmu$s long detection pulses with a repetition rate of 100 kHz. The different curves are obtained by scanning the microwave frequency. The microwave detuning from the atomic transition is $0$ Hz (red curve), $1$ kHz (blue curve), $1.2$ kHz (green curve) and $1.8$ kHz (black curve). The colors in (b) and (c) refer to the same microwave detuning.
 (b) Simulation of the expected signal during the evolution of the atomic sample. (c) Trajectories of the atomic state on the Bloch sphere. (d) Amplitude of the oscillation of the signal during the intermediate $\pi$ pulse. The red dots are experimental data and the black dashed line is the expected amplitude.}
  \label{fig:coherence}
\end{figure}

The atoms are initially prepared in the $\left|F=1,m_{\rm F}=0\right\rangle$ state and probed in the configuration of figure \ref{fig:Rabi}(a). The interferometer is a $\frac{\pi}{2}-\pi- \frac{\pi}{2}$ spin-echo sequence.
The whole sequence was run in less than 500 $\upmu$s to avoid loss of signal due to the atomic cloud expansion and free fall. The $\pi$ pulse has a duration of $74.5~\upmu$s.
The intermediate $\pi$ pulse maps the interferometric phase on the measured observable $\widehat{J_z}+\widehat{N}/2$, which is the projection of the atomic spin vector on the z-axis in the Bloch representation of figure \ref{fig:coherence}(c). To modify the interferometric phase, the frequency of the microwave driving field is scanned with respect to the ground state hyperfine transition. 

Figure \ref{fig:coherence}(a) shows the traces of the atomic state interferometer acquired in real time for different microwave detunings. The result is well understood in terms of rotations of the pseudo-spin vector on the Bloch sphere (figure \ref{fig:coherence}(b)-\ref{fig:coherence}(c)). The amplitudes of the signal oscillation during the intermediate $\pi$ pulse were extracted from the traces and compare well with the predicted ones (figure \ref{fig:coherence}(d)). The characterization shows that the superposition state of an interferometer can be measured without necessarily blurring the fringes. This indicates that the measurement was realized in the weak measurement regime, where the atomic state is not strongly projected. Moreover, in a usual interferometric sequence, the atoms are lost after the destructive measurement. With such a non-demolition measurement at the end of the sequence, the atoms can be reused so as to increase the duty cycle of atomic interferometers \cite{Lodewyck2009}. 

\subsection{Cavity-enhanced measurements}

\label{sec:cavity_enhanced}
As shown in \cite{Vanderbruggen2010,Oblak2005} the optimal level of squeezing $\xi^2=1/(1+\kappa^2)$ is related to the signal to noise ratio $\kappa$ of the measurement:
\begin{equation}
\kappa^2=\frac{\phi^2 N_{\rm at} N_{\rm s}}{2}\propto \rho_0 \eta,
\end{equation}
where $\phi$ is the single atom induced phase shift, $\rho_0$ is the resonant optical density, and $\eta$ is the single atom spontaneous emission probability for a pulse of $N_{\rm s}$ photons.
The coherence loss due to spontaneous emission cannot be avoided since $\eta$ enters in the equation of $\kappa$. Nevertheless, using a cavity to strengthen the atom-light interaction increases the coupling $\phi$ by the cavity finesse $\mathcal{F}$. This results in a SNR enhancement proportional to  $\sqrt{\mathcal{F}}$ \cite{Lye2003} for a given spontaneous emission rate ($\eta\propto N_{\rm s}$).

\begin{figure}[H]
\center
\includegraphics[width= 13 cm]{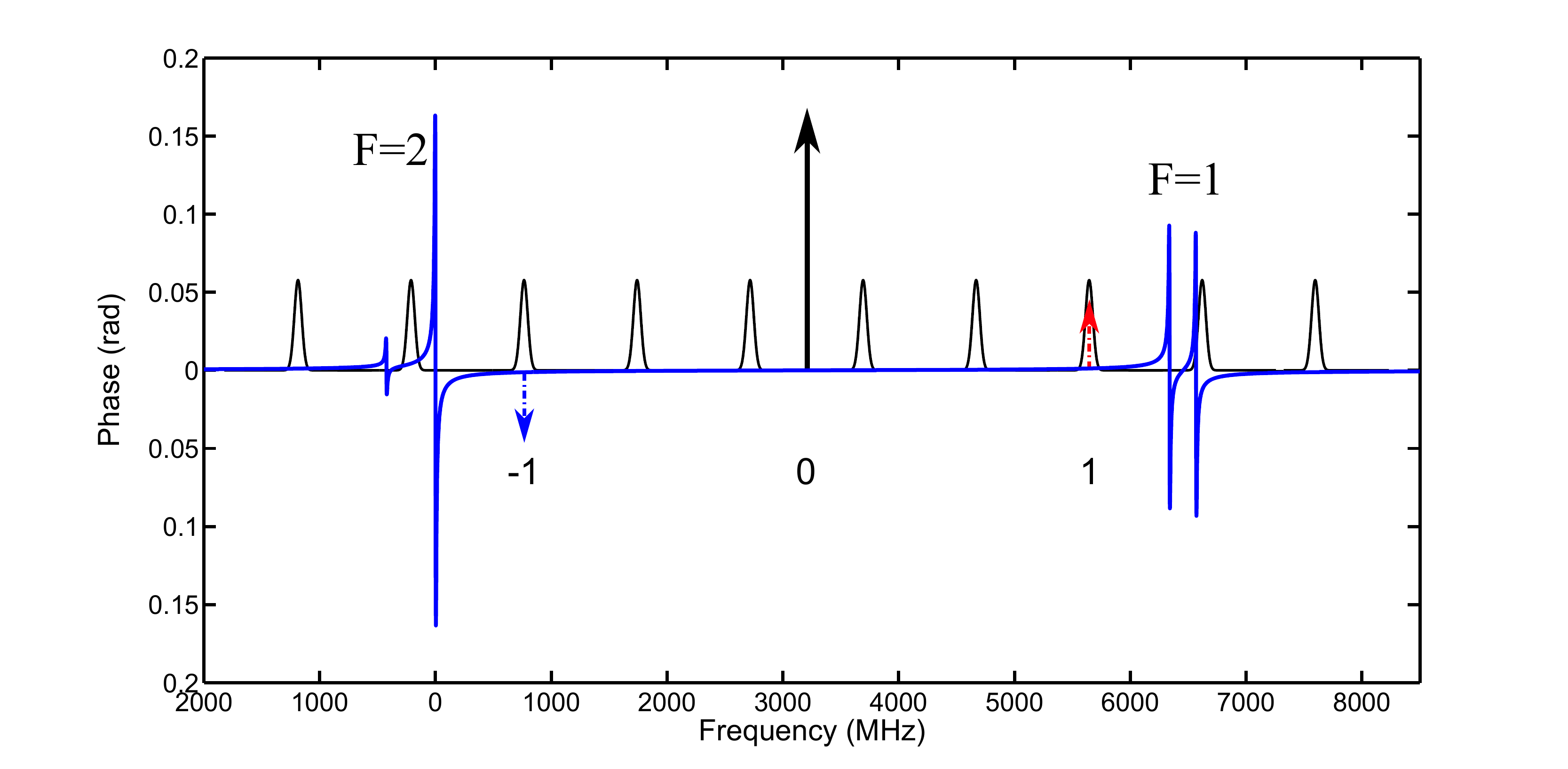}
\caption{Scheme of the Pound-Drever-Hall like measurement method. The carrier is locked at the center of two cavity resonances to suppress cavity noise. Each sideband probes one of the hyperfine states.}
\label{fig:cavity_lines}
\end{figure}

To prevent for the light-shift effect of the carrier, the measurement can be realized in reflection \cite{Teper08}. In the scheme presented in figure \ref{fig:cavity_lines}, to measure the population difference,  the detection realized with the two sidebands is shot noise limited because of the strong local oscillator. In addition, the effective interaction between atoms and photons can take advantage of the trapping and probing that exploits the same cavity mode.

\section{Conclusions}

We have presented an apparatus designed for the production of spin-squeezed cold atomic samples in a dual-frequency high-finesse cavity. We have shown that, thanks to the intra-cavity power enhancement, the chosen configuration is suitable for all-optical trapping of atoms using standard optical telecommunication technologies. 

We described a new non-demolition heterodyne detection scheme. This scheme is limited by the optical shot noise of the probe and presents a high rejection of optical path length fluctuations. We demonstrated the method by performing nondestructive measurements of the coherent evolution of internal atomic states. From the nondestructive measurement of Rabi oscillations, we have shown that the decoherence induced by the probe can be limited by the inhomogeneous light shift from the carrier and not by the spontaneous emission induced by the sideband. We also presented a real-time measurement of the atomic state evolution in a Ramsey interferometer with low decoherence.

Our experiment is particularly suited for metrology with a high squeezing level of large particles numbers. Because of the power build-up at 1560 nm, we will achieve large degenerate samples with low intensity and heterodyne detection can be enhanced by using the cavity at 780 nm. Moreover, using the cavity enhanced atom-light coupling combined with the intrinsic atomic and optical mode overlap should allow us to reach the ultimate sensitivity of the heterodyne measurement device while avoiding inhomogeneous light-shift from the carrier. 

\section{Acknowledgment}
We thank David Holleville, Thierry Botter and Remi Geiger for their participation in the early stage of the experiment. 
This work was supported by IFRAF, DGA, European Union (with STREP program FINAQS), and ESF (EUROQUASAR program). A. B. acknowledges support from EU under an IEF Grant. S.B. acknowledges the Fulbright foundation for support.

\bibliographystyle{unsrt}

\end{document}